\documentclass{jpp}
\usepackage[dvipsnames]{xcolor}
\usepackage{comment}
\usepackage{graphicx}
\usepackage{wrapfig}
\usepackage{subcaption}
\usepackage{caption}
\usepackage[utf8]{inputenc}
\usepackage[T1]{fontenc}
\usepackage{amsmath}
\usepackage{endnotes}
\usepackage{hyperref}

\newtheorem{proposition}{Proposition}

\newtheorem{remark}{Remark}
\let\footnote=\endnote

\def\d{\text{d}}

\shorttitle{Intergenerational Risk Sharing in a Defined Contribution Pension System}

\shortauthor{A. Chen, M. Kanagawa and F. Zhang}

\newcommand{\tmu}{\tilde{\mu}}
\newcommand{\tsigma}{\tilde{\sigma}}

\title{Intergenerational Risk Sharing in a Defined Contribution Pension System: Analysis with Bayesian Optimization}

\author{An Chen\aff{1}\aunote{an.chen@uni-ulm.de},
	Motonobu Kanagawa\aff{2}\aunote{motonobu.kanagawa@eurecom.fr}
	\and Fangyuan Zhang\aff{2}\aunote{fangyuan.zhang@eurecom.fr (Corresponding)}}

\affiliation{\aff{1}Institute of Insurance Science, University of Ulm, Germany
	\aff{2}Data Science Department, Eurecom, France}

\begin{document}
	
	\maketitle
	
	\begin{abstract}
		\begin{center}
			\textbf{Abstract}\\
		\end{center}

		We study a fully funded, collective defined-contribution (DC) pension system with multiple overlapping generations. We investigate whether the welfare of participants can be improved by intergenerational risk sharing (IRS) implemented with a realistic investment strategy (e.g., no borrowing) and without an outside entity (e.g., share holders) that helps finance the pension fund. To implement IRS, the pension system uses an automatic adjustment rule for the indexation of individual accounts, which adapts to the notional funding ratio of the pension system. The pension system has two parameters that determine the investment strategy and the strength of the adjustment rule, which are optimized by expected utility maximization using Bayesian optimization. 
		The volatility of the retirement benefits and that of the funding ratio are analyzed, and  it is shown that  the trade-off between them can be controlled by the optimal adjustment parameter to attain IRS. 
Compared with the optimal individual DC benchmark using the life-cycle strategy, the studied pension system with IRS is shown to improve the welfare of risk-averse participants, when the financial market is volatile.
  

	\end{abstract}

	\section{Introduction}

	Defined Contribution (DC) pension plans constitute key part of pension systems in many countries, such as the ``401k plan'' in the United States and the ``personal pensions'' in the United Kingdom.  In a DC plan, each participant owns her account, pays fixed contributions to the account regularly, and the accumulated contributions are invested in a financial market.  The amount of retirement benefits is determined by the market value of the individual account at the retirement date. By design, DC plans have advantages over traditional pension schemes such as Pay-As-You-Go (PAYG) and Defined Benefit (DB) pension plans in terms of the transparency,  fairness, portability and sustainability, which encourage the prevalence of DC plans.

	However,  DC plans have major issues arising from that each participant bears the investment risk by herself.  One issue is that the participant may not choose a good investment strategy due to the lack of financial expertise. Indeed, a typical DC plan participant in reality tends to perform the naive ``$1/n$ diversification,'' equally dividing her contributions into the default options provided by the plan \citep{benartzi2001naive}.
	Optimal investment strategies studied in the literature  \citep[e.g.,][]{merton1975optimum,cairns1996continuous,cairns2006stochastic,boulier2001optimal,vigna2001optimal,chen2015optimal,menoncin2017mean} can thus not be easily employed by such participants.
	
	Another major issue of DC plans is the incapability of {\em intergenerational risk sharing (IRS)}, which is to diversify non-diversifiable risks within one generation (e.g., those caused by economic shocks) across different generations.  
	For example, a DC plan participant whose accumulation period overlaps an economic depression faces a financial risk that cannot be diversified by herself; even if she can choose an optimal utility-maximizing investment strategy, she may not accumulate enough wealth for retirement. IRS enables diversifying such non-diversifiable risks by sharing the risks among different generations.  By definition, however, IRS requires the pension scheme to be {\em collective}, and thus individual DC plans cannot implement IRS. 
	
	It is well known that carefully-designed collective schemes can implement IRS to improve the welfare of participants  \citep[e.g.,][]{gordon1988intergenerational,allen1997financial,shiller1999social}.
	\citet{gollier2008intergenerational} shows that a collective DC pension plan can improve the welfare of participants compared with an individual benchmark using the optimal life-cycle investment strategy.  Similarly, \cite{cui2011intergenerational} study a collective DB-based hybrid pension scheme where both contributions and pension benefits may be adjusted, and show that it is welfare-improving compared with an optimal individual benchmark.  \cite{chen2016intergenerational} consider a three-pillar pension system in which the second pillar is a collective hybrid plan, and show that there is welfare improvement compared with a corresponding individual DC benchmark.  See \citet[Chapter 7]{barr2008reforming} and \citet{Beetsma16IRS} for an overview of IRS and further references.

	While the seminal work of \cite{gollier2008intergenerational} shows that IRS in a collective DC pension system is welfare-improving,  both his first-best and second-best strategies depend on rather strong assumptions. The first-best strategy attains IRS by enhancing the risk-taking ability of the pension fund, by treating the net present value of  the contributions from all the future generations as part of the fund's {\em total} wealth. Consequently, the fund can invest more than the fund's {\em actual} wealth, i.e., the fund can perform {\em borrowing} for investment, similar to the life-cycle investment strategy for an individual investor \citep{merton1975optimum}. However, borrowing is not realistic for pension funds in reality.  On the other hand, his second-best strategy does not allow  borrowing, but assumes the existence of an outside entity (shareholders) that helps finance the fund. Therefore, it is not clear whether the welfare-improvement by the second-best strategy can be attained without the shareholders.
	
	Given these  limitations of Gollier's analyses, one may ask: {\em Can the welfare-improvement by IRS be attained by a fully funded, collective DC pension system with a realistic investment strategy (e.g., no borrowing) and without an outside entity that helps finance the pension fund?}\   Previous related works do not exactly answer this question, as discussed later in detail. For example, while \citet{chen2016intergenerational} show that their collective scheme with IRS is welfare-improving compared with the corresponding individual DC scheme, it is assumed that both the collective and individual schemes use the {\em same} investment strategy;  therefore, it is not  clear whether their IRS is welfare-improving as compared with the {\em optimal} individual investment strategy.
	
	Our main aim is to investigate the above question.
	To this end, we consider a stylized model for a fully funded, collective DC pension fund with multiple overlapping generations, which we call the {\em IRS-DC} model. As \cite{gollier2008intergenerational}, each participant  pays a fixed annual contribution to the pension fund, and the fund makes investment on behalf of the participants.  Different from the first-best strategy of \cite{gollier2008intergenerational}, however, the fund is not allowed to perform borrowing. Each participant has her own account in the fund, which accumulates her contributions and is indexed to the fund's investment performance; the account value at the retirement date determines her pension benefit. The indexation rate of individual accounts is automatically adjusted to the (notional) funding ratio of the pension fund by the adjustment rule of  \citet{goecke2013pension}.   This automatic adjustment rule is the device for implementing IRS in our model. In contrast to the second-best strategy of \citet{gollier2008intergenerational}, the fund is fully-funded and does not rely on any external entity to implement IRS.
	
	We analyze how the automatic adjustment rule stabilizes the funding ratio and the benefits of participants to attain IRS. Analytic expressions are derived for the funding ratio and benefits. It is shown that there is a trade-off between the stability of the funding ratio and that of benefits, and that this trade-off is controlled by the strength of the automatic adjustment rule. That is, benefits can be made more stable by increasing the volatility of the funding ratio, and vice versa.  IRS can be attained by balancing this trade-off. 
	
	Automatic adjustment rules in pension systems have been not only studied in the literature \citep[e.g.,][]{cui2011intergenerational,chen2016intergenerational,bams2016optimal,donnelly2017discussion} but also  applied to real pension systems in such countries as Sweden and the Netherlands  \citep[Chapter 2]{paris2021pensions}. They are used for improving the sustainability of a pension fund and for providing stable benefits to participants  \citep[e.g.,][]{settergren2001automatic,barr2011improving}. 
	However, a formal analysis is missing for justifying such use of automatic adjustment rules in collective pension systems.
	Our analysis thus provides a first step in this regard.
	
	The IRS-DC model has two parameters, one for the investment strategy and the other for the automatic adjustment rule. For optimizing these parameters, we define an expected utility maximization problem that involves the benefits of all the generations including those in the future, following \cite{gollier2008intergenerational}. As this optimization problem cannot be solved analytically, we solve it numerically using {\em Bayesian optimization}, a machine learning approach to optimizing a black-box function \citep[e.g.,][]{shahriari2016taking}. As discussed later, the use of Bayesian optimization is our computational contribution, in line with the recent deployments of machine learning in the insurance literature \citep[e.g.,][]{hainaut2018neural,gabrielli2020neural,wuthrich2020bias,scognamiglio2021calibrating,schnurch2022point}.
	
To answer the question above, our main finding is that IRS can improve the welfare of participants without borrowing and shareholders, {\em if} the financial market is volatile and the participants are risk-averse; IRS may not be welfare-improving if this condition is not satisfied.
	We compare the welfare of the IRS-DC plan participants and the welfare of the corresponding individual DC plan participants, where the latter uses the optimal life-cycle strategy \citep{merton1975optimum}.
	Several different settings of the financial market and the risk aversion of participants are investigated, and the above finding is obtained.

	The paper proceeds as follows. Section \ref{sec:model} introduces the IRS-DC pension model, which is analyzed in Section \ref{sec:analysis-section}.
	Section \ref{sec:BO} explains the expected utility maximization problem, how to solve it with Bayesian optimization, and the setup for simulations.
	Section \ref{sec:numerical-analysis} presents numerical analyses, including the funding ratio process, the individual benefit accounts of the IRS-DC fund, and the certainty equivalents of the participants.
 Section \ref{sec:conclusion} concludes.  
 \textcolor{black}{
 The appendix contains a short tutorial on Bayesian optimization, the proofs of analytic results, and additional numerical analyses. }


	\section{Pension Model} \label{sec:model}
	
	This section describes the IRS-DC pension model. 
	%
	The pension fund contains multiple overlapping generations, where there are always incoming and outgoing generations. Each generation pays fixed contributions annually to the pension fund. 
	Before explaining the details, we summarize below the key features of the pension fund: 
	\begin{enumerate}[a)]
		\item The pension fund collectively invests the contributions from different generalizations (participants) in a financial market.  
		
		\item Each participant maintains her account in the pension fund that records her accumulated pension rights.  
		
		\item The growth rate of individual accounts is automatically adjusted based on the fund's investment performance and a notional funding ratio, so that intergenerational risk sharing is implemented.
		
	\end{enumerate}
	
	
	Section \ref{finan} describes the IRS-DC pension model in detail.
	Section \ref{sec:comparison-related-models} compares it with related pension models in the literature.
	
	\subsection{Description of the IRS-DC Pension Model}\label{finan}

	\begin{figure}
		\centering
		\includegraphics[width=0.6\linewidth]{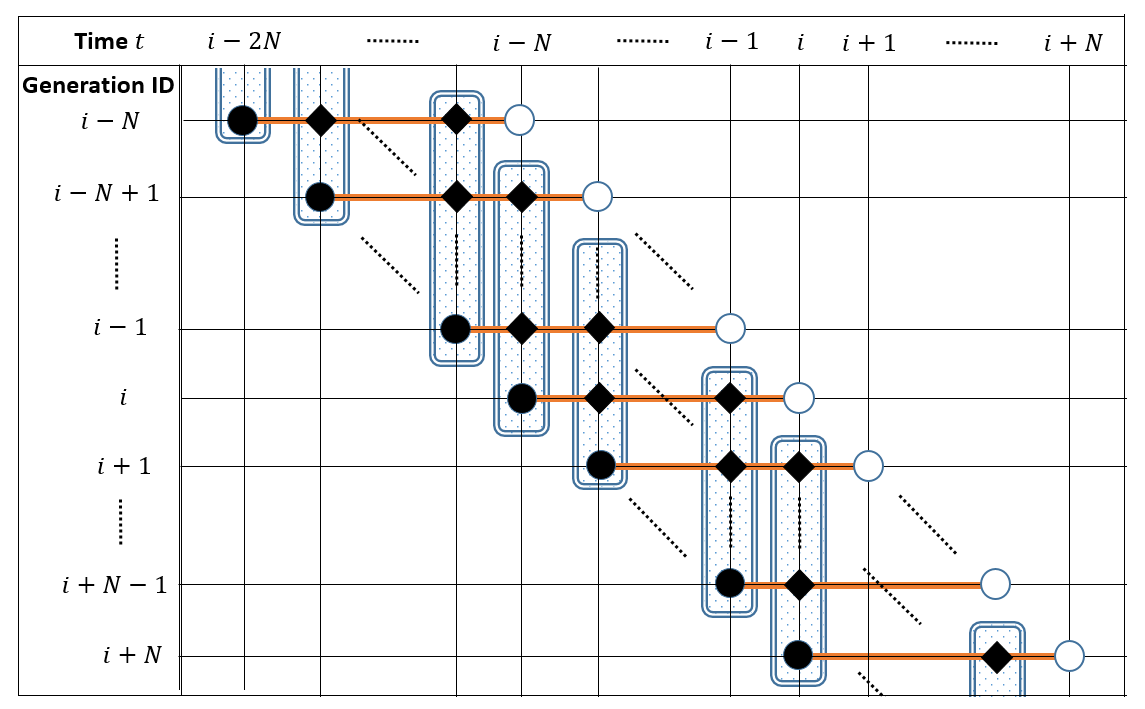}
		\caption{Schematic illustration of the pension model. The horizontal axis indicates time $t$, and the vertical axis generation identifiers. For each generation, the black and white circles indicate the time points when that generation joins and retires from the pension fund, respectively; the orange line indicates the duration of the participation in the fund. For example, the generation $i$  joins the fund at time $t = i-N$ and retires at $t = i$.  For each time point, the blue box indicates the current participants of the pension fund, and black diamonds indicate those participants (only shown for the generations displayed in the figure). For example, at time $t = i-1$, the generations $i, i + 1, \dots, i+N-1$ are participating in the fund. }
		\label{fig:illust-pension}
	\end{figure}

	
	Figure \ref{fig:illust-pension} provides a schematic illustration of our pension model.
	The pension fund covers $N$ overlapping working generations in each operating year (e.g., $N = 40$). The pension fund is fully-funded.  For simplicity, we assume that each generation consists of one hypothetical participant.
	Let $t \geq 0$ denote the time, with the unit being one year. 
	We assume that the fund starts at time $t = 0$ with $N$ initial generations. 
	
	\subsubsection{Generation Identifier}
	We use an integer $i \in \mathbb{N}$ as the identity of the generation who retires at time $t = i$ (see Figure \ref{fig:illust-pension}). 
	Namely, the generation $i$ joins the fund at time $t = i- N$ and leaves the fund at time $t = i$; thus, this generation is in the fund for $N$ years. 
	Using this notation, we can define the set of all the working generations in the fund at any time point $t \geq 0$ as 
	\begin{equation*}
		I_{w}(t):=\{i\vert i= [t] + 1,\ [t] + 2,\ \dots,\ [t] + N \},
	\end{equation*}
	where $[t]$ denotes the integer part of $t$ (e.g., if $t = 3.4$ then $[t] = 3$; if $t = 3$ then $[t] = 3$).
	
	\subsubsection{Financial Market and Pension Asset Dynamics} 
	
	\label{sec:financial-market}
	
	We consider a  financial market where there exist two investment opportunities: a risky asset (e.g., a stock) and a risk-free asset (e.g., a bank account or a bond), denoted by $S(t)$ and  $F(t)$, respectively.  
	Specifically, we consider the Black-Scholes market, where $S(t)$ is driven by a diffusion process with constant drift $\mu > 0$ and volatility $\sigma > 0$, while $F(t)$ develops at a risk-free rate $r > 0$ such that $\mu > r$:  
	\begin{align}
		&\d S(t)=\mu S(t)\d t+\sigma S(t)\d Z(t),\quad S(0)=1;\label{eq:stock}\\
		&\d F(t)=r F(t)\d t, \quad F(0)=1,\label{eq:bank}
	\end{align}
	where $Z(t)$ is a standard Brownian motion under the real-world probability measure. 
	
	
	Let $A(t)$ denote the asset of the pension fund at time $t$, with $A(0) > 0$ being the initial asset.  For simplicity, we assume that the fund invests a constant fraction $\pi \in [0,1]$ of the pension asset $A(t)$ in the stock $S(t)$ and the rest $1-\pi$ in the risk-free asset $F(t)$; thus, we can write the dynamics of the pension asset as
	\begin{equation} \label{eq:330}
		\d A(t)=\frac{\pi A(t)}{S(t)}\d S(t)+\frac{(1-\pi)A(t)}{F(t)}\d F(t), \quad\pi\in[0,1].
	\end{equation}
	We assume that the fund is prohibited from  from borrowing ($\pi>1$) and short selling ($\pi<0$).
	
	At the beginning of each year $t = 0, 1, 2 \dots$, each working generation pays a constant amount of contribution, $c > 0$, to the pension fund. Since there are $N$ working generations, the fund thus receives a total of $Nc$ contributions at the beginning of each year. At the same time, the fund pays a lump-sum benefit to the generation $t$, who retires at time $t$.
 
 The dynamics of the pension asset can thus be written as
	\begin{align}
		&\d A(t)= A(t)(\pi(\mu-r)+r)~\d t+ A(t)\pi\sigma~\d Z(t), \quad A(0)=A_0,\quad t>0,\label{eq:asset-345} \\
		&A(t)_+ =  A(t) + N c - B_t(t), \quad t = 0, 1, 2, \cdots, \label{dynamics:asset}
	\end{align}
	where \textcolor{black}{\eqref{eq:asset-345}  is obtained by substituting \eqref{eq:stock} and \eqref{eq:bank} into \eqref{eq:330}}, $A(t)_+ := \lim_{\varepsilon \to +0} A(t+\varepsilon)$ denotes the right continuous limit, and $B_t(t)$ is the benefit paid to the generation $t$ who has just retired and defined in   \eqref{eq:374} and \eqref{dynamic:benefit} below (the double $t$  notation of $B_t(t)$ is deliberate and its meaning will be clear shortly).
	
	Therefore, the pension asset develops continuously over time $t >0$, while there is a jump at each integer time $t = 0, 1, 2, \cdots$ (i.e., at the beginning of each year) when there are incoming and outgoing cash flows of $Nc$ and $B_t(t)$, respectively.  
	
	\subsubsection{Individual Accounts and Retirement Benefits} 
	
	Like pure DC and notional DC plans, each participant (generation) in our IRS-DC pension fund has her individual account, which keeps track of her pension rights.
	It records her annual contributions and grows according to the indexation rate \eqref{benefit-growth-rate} defined below.
	The terminal value of the account at the time of retirement becomes the lump-sum retirement benefit.

	More formally, let $B_i(t)$ denote the individual account of generation $i \in \mathbb{N}$ at time $t \in [i-N, i]$, which starts from $B_i(i-N) = 0$ when this generation enters the fund at time $t = i-N$. 
	Then we define its dynamics as 
	\begin{align}
		&	B_i(t)_+= B_i(t) + c, \quad \quad 	t = i - N,\ \   i - N + 1,\ \dots,\ \ i - 1, \label{eq:374} \\
		& 	\d B_i(t)=B_i(t)g(t)\d t, \quad \quad  i - N \leq t \leq i \label{dynamic:benefit}.
	\end{align}
	where $g(t)$ is the indexation rate at time $t$, defined in \eqref{benefit-growth-rate} below. 
	Namely, the individual account $B_i(t)$ grows continuously from the entry time $t = i - N$ until retirement at $t = i$ according to the indexation rate $g(t)$ as \eqref{dynamic:benefit}, while the account accumulates the annual contribution $c$ at the beginning of every year as \eqref{eq:374}.   
	The account value at the time of retirement $t = i$, i.e, $B_t(t)$, is the retirement benefit of the generation $i$; see \eqref{dynamics:asset}.
	\subsubsection{Indexation Rate and Notional Liability}
	We now define the indexation rate $g(t)$ that determines the growth rate of individual accounts as in \eqref{dynamic:benefit}, which in turn affects the pension's asset dynamics in \eqref{dynamics:asset}. 
	To this end, following  \citet{bams2016optimal} and \citet{donnelly2017discussion}, we first define a {\em notional liability} of the fund as
	\begin{equation} \label{eq:liability}
		L(t) := \sum_{i \in I_w(t)} B_i(t), \quad t \geq 0.
	\end{equation}
	That is, we define the notional liability $L(t)$ of the fund at time $t$ as the sum of individual accounts $B_i(t)$ for the current working generations $i \in I_w(t)$.

	
	If there is no cash flow in \eqref{dynamics:asset}, the solution to the asset process  \eqref{eq:asset-345} \textcolor{black}{is given by stochastic exponential, which is a classic result in financial mathematics (e.g., \citealt[Chapter 5]{karatzas1991brownian}), as}  
	\begin{equation}   \label{eq:asset-solution-413}
		A(t)  = A(0)  \exp \left( \int_{0}^t \tilde{\mu}  \d s + \int_{0}^t   \tilde{\sigma}  \d Z(s) \right), 
	\end{equation}
	where $\tilde{\mu} > 0$ and $\tilde{\sigma} > 0 $ are constants defined as 
	\begin{equation} \label{eq:expected-ann-log-return}
		\tilde{\mu} :=\pi( \mu -r) + r - \frac{1}{2} \pi^2 \sigma^2, \quad \quad   \tilde{\sigma} := \pi \sigma .
	\end{equation}
	Following \citet[Eq.~(5)]{goecke2013pension},\footnote{Our indexation rate corresponds to \citet[Eq.~(5)]{goecke2013pension} with $\rho_{\rm target} = 0$. } we then define the indexation rate $g(t)$ as 
	\begin{equation} \label{benefit-growth-rate}
		g(t) = \tilde{\mu}  +\theta\ln\left(\frac{A(t)}{L(t)}\right), \quad t \geq 0,
	\end{equation}
	where  $\theta > 0$ is a constant,  $\tilde{\mu}$ is from \eqref{eq:expected-ann-log-return}, $A(t)$ is the pension fund's asset process \eqref{eq:asset-345} \eqref{dynamics:asset}, and $L(t)$ is the notional liability \eqref{eq:liability}.  
 
	
		\begin{figure}
		\centering
		\includegraphics[width=0.6\linewidth]{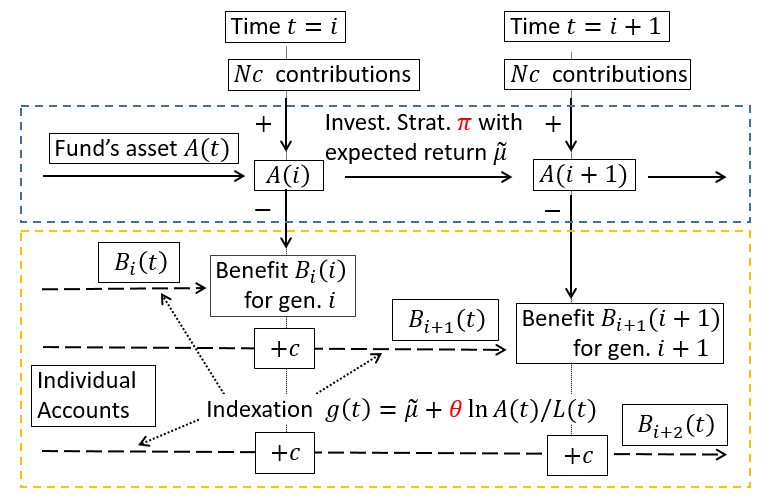}
		\caption{Schematic illustration of the cash flows in our pension model for two time points $t = i$ and $t = i + 1$. In the blue dotted box, $A(t)$ denotes the pension fund's asset, whose dynamics follow \eqref{eq:asset-345} and \eqref{dynamics:asset}. In the orange dotted box, $B_i(t)$, $B_{i+1}(t)$, and $B_{i+2}(t)$ denote the individual accounts of generations $i$, $i+1$, and $i+2$, respectively, whose dynamics follow \eqref{eq:374} and \eqref{dynamic:benefit}.
			At time $t = i$, there is an incoming cash flow of $Nc$ contributions from the $N$ working generations, and an outgoing cash flow of the retirement benefit $B_i(i)$ for the generation $i$, who retires at time $t = i$; see \eqref{dynamics:asset}.  This retirement benefit is determined by the individual account $B_i(t)$, which accumulates the annual contribution $c$ at the beginning of each year and develops continuously according to the indexation rate $g(t) = \tilde{\mu} + \theta \ln( A(t)/L(t) )$; see \eqref{dynamic:benefit}. On the other hand, the asset process $A(t)$ develops continuously according to the investment performance of the portfolio with the constant-mix strategy $\pi$; see \eqref{eq:asset-345}. Each ``$+c$'' represents the accumulation of the annual contribution in the individual account: At time $t$, the amount $c$ is added to each of $B_{i+1}(t)$ and $B_{i+2}(t)$. 
		}
		\label{fig:cash-flow}
	\end{figure}
{\color{black}{Figure \ref{fig:cash-flow} provides a schematic illustration of the cash flows in our pension model for two time points $t = i$ and $t = i + 1$. In the following, let us take a closer look at the role of various important factors. }}
	
	\subsubsection{The Role of the Indexation Rate}
	\label{sec:disccussion-indexation-rate}
	
	While we will present a more formal analysis in Section \ref{sec:analysis}, we provide here an intuitive discussion of how the indexation rate \eqref{benefit-growth-rate} works.
	As defined in \eqref{dynamic:benefit}, the indexation rate $g(t)$ controls the growth rate of individual accounts $B_i(t)$.
	The first term $\tilde{\mu}$ in \eqref{benefit-growth-rate} is the expected  annual log return using the same investment strategy $\pi$ in the financial market {\em without} participating in the pension fund. In the second term, $A(t) / L(t)$ is the {\em notional funding ratio} that quantifies the balance between the asset $A(t)$ and the notional liability $L(t)$. 
	The second term adjusts the growth rate of individual accounts $B_i(t)$, and the parameter $\theta \geq 0$ specifies the strength of the adjustment.
	

\textcolor{black}{	If $\theta = 0$, then individual accounts grow {\em deterministically} at the rate $\tilde{\mu}$; therefore in this case, the retirement benefits are {\em ex-ante determined}, i.e., the pension plan becomes a defined-benefit plan, thus removing the investment risk of pension participants.
	However, it risks the sustainability of the pension fund, as the fund is of a defined-contribution type by design, and thus there is no way of adjusting the contributions when the fund is underfunding. }
	
\textcolor{black}{	On the other hand, if one sets a large value of $\theta > 0$, then pension participants bear more investment risks to improve the sustainability of the pension fund. 
	For example, suppose that the pension asset exceeds the notional liability, i.e., $A(t) > L(t) = \sum_{i \in I_w(t)}B_{i}(t)$. One can interpret this situation as that the fund  yields a high return in the investment and thus there is a ``surplus.''
	Then the log notional funding ratio becomes positive, $\ln (A(t) / L(t)) > 0$, and the indexation rate $g(t)$ \textcolor{black}{shall be larger than $\tilde{\mu}$}; therefore individual accounts $B_i(t)$ grow faster, reflecting the high investment return. 
	On the other hand, if $A(t) < L(t) = \sum_{i \in I_w(t)}B_{i}(t)$, which happens when the fund yields a low return and thus there is a ``deficit.'' In this case, we have $\ln (A(t) / L(t)) < 0$ and thus the indexation rate $g(t)$ \textcolor{black}{shall be smaller than $\tilde{\mu}$}; therefore individual accounts $B_i(t)$ grow more slowly, reflecting the low investment return.  }
	
	\textcolor{black}{
	This argument implies that the adjustment parameter $\theta$ should be neither too small nor too large. One should choose $\theta$ appropriately to achieve a good trade-off between the risks of individual participants and the pension fund. 
	We present a more formal analysis in Section \ref{sec:analysis}.
 }

	\subsection{Comparison with Related Pension Models} 
	
	\label{sec:comparison-related-models}
	
	We compare the IRS-DC model with related pension models in the literature.
	\citet{goecke2013pension} studies the indexation rate \eqref{benefit-growth-rate} for the return smoothing in a self-financing pension plan. \citet{goecke2013pension}'s model consists only of one generation, and there exists no cash flow of contributions and payments. 
	The earlier work by \citet{baumann2008pension} considers the indexation rate \eqref{benefit-growth-rate} where the risk-free rate $r$ is used instead of  $\tilde{\mu}$. 
	Our model is a continuous-time version of the discrete-time model of \citet{bams2016optimal}, which itself is an extension of the overlapping generations model of  \citet{gollier2008intergenerational}. \citet{bams2016optimal} use the indexation rate in \eqref{benefit-growth-rate}, but do not analytically study its use. 
	\citet{donnelly2017discussion} considers a funded collective DC pension plan. \citet{donnelly2017discussion}'s model consists of {\em fixed} multiple overlapping generations, and there exists no new incoming generation. \citet{donnelly2017discussion} uses an automatic adjustment rule similar to \citet{goecke2013pension}'s (and thus ours) but is different in its concrete form.

	\citet{cui2011intergenerational} consider a funded DB-based hybrid pension system that can adjust both benefits and contributions. Their pension model does not have individual accounts. The present value of base benefits and contributions are made equivalent ex-ante but there is no direct link between one's actual benefits and contributions. Their model is DB-based in this sense. They use automatic adjustment rules for contributions and retirement benefits based on the funding ratio. \citet{chen2016intergenerational} consider a hybrid pension plan as their second-pillar pension system, in which there exist individual accounts. They also use automatic adjustment rules for contributions and individual accounts' indexation rates. While the adjustment rules of \citet{cui2011intergenerational} and \citet{chen2016intergenerational} are conceptually similar to ours, they are different in their forms. For example, \citet{chen2016intergenerational} use the ``tangent hyperbolic adjustment function,'' while our adjustment rule is based on the log notional funding ratio. Moreover, \citet{cui2011intergenerational} and \citet{chen2016intergenerational} define the liabilities in a DB manner, taking into account future retirement benefits, while we define our notional liability in a DC manner, i.e., as the sum of current individual account values. Again, our liability is notional since the fund does not provide any promise on retirement benefits.

	
	Automatic adjustment mechanisms have been implemented in real pension systems; see  \citet[Chapter 2]{paris2021pensions} for an overview. Notably, Sweden's first-pillar notional DC pension system uses an automatic adjustment rule for the indexation rate of individual accounts \citep{settergren2001automatic}.
	This adjustment mechanism is based on a notional funding ratio\footnote{The funding ratio in Sweden's notional DC pension system is notional since both ``assets'' and ``liabilities'' are notional, as it is a PAYG system.} and is conceptually similar to other rules discussed here 
	\citep[See e.g.,][Eqs.\ (6.2) and (6.3)]{Hagen13-SwedishPension}.
	The Swedish first-pillar notional DC pension system defines its notional liability essentially in the same way as \eqref{eq:liability} \citep[Eq.~(3)]{settergren2001automatic}.\footnote{Since retirees leave the fund immediately after receiving lump-sum benefits, Eq.~(5) in  \citet{settergren2001automatic} does not exist in our case and thus Eq.~(3) in \citet{settergren2001automatic} is equal to the notional liability in \eqref{eq:liability}.}


	\section{Analysis }
	\label{sec:analysis-section}
	
	\label{sec:analysis}
	
	We present an analysis of the IRS-DC pension model in Section \ref{sec:model}, focusing on the role of the indexation rate \eqref{benefit-growth-rate} for achieving IRS.  
	In particular, we study how the adjustment parameter $\theta$ in the indexation rate impacts the funding ratio, which measures the stability of the pension fund, and the retirement benefits of individual participants. 
	
	
	In Section \ref{sec:dynamics-fund-ratio}, we first study the dynamics of the log funding ratio.
	Based on this, we  analyze the effects of the adjustment parameter on the dynamics of the funding ratio in Section \ref{sec:effects-adjust-fund-ratio}, and on the retirement benefit  of an individual participant in Section \ref{sec:effects-adjust-pension-benefits}.
	In the latter, we obtain an analytic expression of the retirement benefit in terms of the log funding ratio and the adjustment parameter. 
	Based on this expression, we compare the retirement benefits of the IRS-DC plan and the corresponding pure DC plan in Section \ref{sec:comparison-pure-DC}.
	This last analysis provides insights into how IRS works in the IRS-DC plan.
	
	\subsection{Dynamics of the Log Funding Ratio} 
	
	\label{sec:dynamics-fund-ratio}

	We start by analyzing the dynamics of the {\em log funding ratio} defined as
	\begin{equation} \label{eq:log-funding-ratio-268}
		\rho(t) := \ln \left( \frac{A(t)}{L(t)} \right) .
	\end{equation}
	\citet[Proposition A.1]{goecke2013pension} shows that $\rho(t)$ is an Ornstein-Uhlenbeck process under the assumption that there exists no cash flow. 
	\citet[Section 3.1]{baumann2008pension} obtain a similar result, but again assuming no cash flow. 
	Since our model involves explicit cash flows as in \eqref{dynamics:asset}, these earlier results are not directly applicable.  
	Nevertheless, we show here that $\rho(t)$ in our model is also an Ornstein-Uhlenbeck process if the time $t$ is between integer time points. (Recall that cash flows in our model occur only at integer time points; see \eqref{dynamics:asset}).
	This result, and intermediate derivations, are later used for deriving further results, so we present them here for completeness.

	Let $t_0 \in \mathbb{N}$ be an arbitrary integer time point, which corresponds to the beginning of a year.
	Then by \eqref{eq:asset-345}, with \textcolor{black}{$A(t_0)_+$} being the initial value \textcolor{black}{after the contributions}, the asset process $A(t)$ for $t_0 < t < t_0 + 1$ is written as
	\begin{align*}
		A(t) = {\color{black}{A(t_0)_{+}}} \exp\left( \int_{t_0}^t \tilde{\mu} \d s + \int_{t_0}^t \tilde{\sigma} \d Z(s) \right).
	\end{align*} 
	Notice the difference from the previous expression \eqref{eq:asset-solution-413}, which starts from $t = 0$ and holds only under the assumption that there exists no cash flow. 
	Similarly, by \eqref{dynamic:benefit}, \eqref{eq:liability} and \eqref{benefit-growth-rate}, the notional liability $L(t)$ is given as
	\begin{equation} \label{eq:liability353}
		L(t) = {\color{black}{L(t_0)_+}}  \exp \left(  \int_{t_0}^t \tilde{\mu} \d s +  \theta \int_{t_0}^{t}  \rho(s) \d s  \right). 
	\end{equation}
	
	Then for $t_0 < t < t_0 + 1$, the log funding ratio $\rho(t)$ can be expanded as
	\begin{align}
		\rho(t) &= \ln A(t) - \ln L(t) \nonumber \\
		&=   \ln {\color{black}{A(t{_0})_+}} +  \int_{t_0}^t \tilde{\mu}\d s + \int_{t_0}^t \tilde{\sigma}   \d  Z(s)    - \ln {\color{black}{L(t_0)_+}} -  \int_{t_0}^t \tilde{\mu} \d s - \theta \int_{t_0}^t  \rho(s) \d s  \label{eq:fund-ratio-exp369} \\
		& ={\color{black}{ \rho(t_0)_+}}  - \theta \int_{t_0}^t \rho(s)  \d s   +  \int_{t_0}^t  \tilde{\sigma}  \d  Z(s). \label{eq:log-fund-ratio-276}
	\end{align}
	The last expression is obtained because the two identical terms  $\int_{t_0}^t \tilde{\mu}\d s$ in \eqref{eq:fund-ratio-exp369}  are cancelled out. This is the result of the expected log return $\tilde{\mu}$ being used in defining the indexation rate \eqref{benefit-growth-rate}, which in turn results in \eqref{eq:liability353}.
	

	Equation \eqref{eq:log-fund-ratio-276} indicates that the log funding ratio $\rho(t)$ for $t_0 < t < t_0 + 1$ is an Ornstein-Uhlenbeck process with initial value $\rho(t_0)_+$ \citep[e.g.,][p.~358]{karatzas1998brownian}, which can be written as
	\begin{equation} \label{eq:funding-ratio-331}
		\rho(t) = e^{ - \theta (t - t_0) } {\color{black}{ \rho(t_0)_+}} +  \tilde{\sigma} \int_{t_0}^t e^{ -  \theta ( t - s ) } \d Z(s), \quad t_0 \in \mathbb{N}, \quad t_0 < t < t_0 + 1.
	\end{equation}
	This expression shows that $\rho(t) = \ln (A(t) / L(t))$ is mean-reverting in the sense that, irrespective of the value of \textcolor{black}{$\rho(t_0)_+$}, it tends to $0$  (in expectation) as $t$ increases. In other words, the funding ratio $A(t) / L(t)$ tends to $1$ as $t$ increases. 
	
	
	\subsection{Effects of the Adjustment Parameter on the Funding Ratio} 
	\label{sec:effects-adjust-fund-ratio}

	Based on the expression \eqref{eq:funding-ratio-331}, we next study how the adjustment parameter $\theta$ affects the dynamics of the log funding ratio $\rho(t)$. \textcolor{black}{We summarize key observations in the following proposition, the proof of which can be found in Appendix \ref{sec:proof-prop-fund-ratio}.}
	
\textcolor{black}{
\begin{proposition}
     \label{lemma:funding-ratio}
Let $\rho(t) = \ln ( A(t) / L(t) )$  be the log funding ratio and $\theta > 0$ be the adjustment parameter of the indexation rate $g(t)$. Let $t_0 \in \mathbb{N} \cup \{0 \}$ and $t_0 < t < t _0 + 1$. Then we have the following:
	\begin{enumerate}
	 \item 
  The conditional expectation and variance of $\rho(t)$ given $\rho(t_0)_+$ are given by  
	\begin{align}
		\mathbb{E}[ \rho(t) \mid \rho(t_0)_+ ] & = e^{-\theta(t-t_0)} \rho(t_0)_+, \label{eq:lemma-cond-exp-290}  \\
		\mathbb{V}[\rho(t) \mid \rho(t_0)_+] &   =   \tsigma^2     \int_{t_0}^{t}   e^{- 2\theta (t  - s)}    \d s 
  \label{eq:lemma-cond-var-291},
	\end{align} 
 where  $\tilde{\sigma} > 0$ is the standard deviation of the annual log return in \eqref{eq:expected-ann-log-return}.
	\item As $\theta$ tends to $zero$, the conditional expectation of $\rho(t)$ given $\rho(t_0)_+$ tends to $\rho(t_0)_+$: 
	\begin{equation}\label{eq:limit_0_condi_mean}
	     \lim_{\theta\to +0}\mathbb{E}[ \rho(t) \mid \rho(t_0)_+ ]=\rho(t_0)_+,
	\end{equation}
 and the conditional variance of $\rho(t)$ given $\rho(t_0)_+$ tends to $\tilde{\sigma}^2$ times $t-t_0$: 
	   \begin{equation}\label{eq:limit_0_condi_var}
	       \lim_{\theta\to +0}	\mathbb{V}[\rho(t) \mid \rho(t_0)_+]=\tsigma^2 (t-t_0).
	   \end{equation}
	 	 \item 
    As $\theta$ tends to infinity, the conditional expectation and variance of $\rho(t)$ given $\rho(t_0)_+$ tend to zero:
	  \begin{equation}\label{eq:limit_f_condi_mean} 
	           \lim_{\theta\to \infty} \mathbb{E}[ \rho(t) \mid \rho(t_0)_+ ]=0, \quad  \lim_{\theta\to \infty}	\mathbb{V}[\rho(t) \mid \rho(t_0)_+]=0.
	      \end{equation}
	\end{enumerate}
	\end{proposition}}

	Proposition \ref{lemma:funding-ratio} shows how the adjustment parameter $\theta$ affects the notional funding ratio $A(t)/L(t)$ and thus the stability of the pension fund. 
	Point (iii) shows that a larger $\theta$ lets $A(t)/L(t)$ approach $1$ more quickly and thus makes the fund more stable,  while point (ii) indicates that a smaller $\theta$ makes the fund more volatile.
	Recall that the value of $\theta$ determines how strong the adjustment in the indexation rate $g(t)$ works for the individual accounts $B_i(t)$; see \eqref{benefit-growth-rate}.
	Therefore, a larger $\theta$ results in a stronger adjustment of the individual accounts $B_i(t)$, so that  the notional liability  $L(t) = \sum_{i \in I_w(t)} B_i(t)$ is adjusted more quickly to match the fund's asset $A(t)$; this is an intuitive explanation of how a large $\theta$ improves the stability of the pension fund. 
	
	While a larger $\theta$ may be beneficial for the fund's stability, it results in a stronger adjustment of the individual accounts $B_i(t)$, which may make the retirement benefits volatile. 
	Therefore it is important to understand the effects of $\theta$ on the retirement benefits; we analyze this next.

\subsection{Effects of the Adjustment Parameter on the Pension Benefits}
\label{sec:effects-adjust-pension-benefits}

We next study how the adjustment parameter $\theta$ affects the retirement benefit of each generation.
To this end, we obtain an analytic expression of the retirement benefit in terms of the log funding ratio, as summarized in the following proposition.
\textcolor{black}{The proof can be found in Appendix \ref{sec:proof-theo-benefits}.}


\begin{proposition} \label{theo:benefits}
	Let $c > 0$ be the annual contribution,  $\tilde{\mu} >0$ and $\tilde{\sigma} > 0$ be the mean and the standard deviation of the annual log return in \eqref{eq:expected-ann-log-return}, $\rho(t) = \ln (A(t) / L(t))$ be the log funding ratio, and $\theta > 0$ be the adjustment parameter of the indexation rate $g(t)$ in \eqref{benefit-growth-rate}. 
	Then for generation $i \in \mathbb{N}$, the retirement benefit $B_i(i)$ is given by
	\begin{align}
	\nonumber 	& B_i(i) =   
		\\ & c \sum_{n=1}^N  \exp  \bigg(  \underbrace{n \tmu}_{\color{black}{\rm (I)}} +  \underbrace{ (1 - e^{-\theta}) \sum_{\ell=1}^n \rho(i-\ell)_+ }_{\color{black}{\rm (II)}}   + \underbrace{ \tsigma \sum_{\ell = 1}^n \int_{i - \ell}^{i-\ell + 1}  \left( 1 - e^{-\theta (i-\ell+1 - s)} \right) \d Z(s) }_{\color{black}{\rm (III)}} \bigg). \label{eq:CDC-benefit-genN}
	\end{align}
\end{proposition}

Proposition \ref{theo:benefits} enables studying the effects of the adjustment parameter $\theta$ on the retirement benefit $B_i(i)$ of the $i$-th generation, who retires at time $t = i$. 
The expression \eqref{eq:CDC-benefit-genN} consists of $N$ terms, in which each term is indexed by $n = 1,\dots,N$. (Recall that $N$ is the total number of years each generation contributes to the fund).
One can understand the $n$-th term in \eqref{eq:CDC-benefit-genN} as corresponding to the contribution $c$ made at time $t = i - n$, i.e.,   $n$ years before the retirement at time $t = i$.

We can make the following observations for the exponent of the $n$-th term in \eqref{eq:CDC-benefit-genN}: \\
\begin{itemize}
	\item The \textcolor{black}{term (I)} corresponds to the deterministic growth term  $\tilde{\mu}$  in the indexation rate $g(t)$; see \eqref{benefit-growth-rate}.\\
	
	\item The  \textcolor{black}{term (II)} represents the effects of the fund's ``surplus'' or ``deficit'' in the last $n$ years before the retirement. One can understand that there is a ``surplus'' if $\sum_{\ell = 1}^n \rho(i-\ell)_+ > 0 $; in this case the retirement benefit increases accordingly, as a redistribution of the surplus. On the other hand, there is a ``deficit'' if  $\sum_{\ell = 1}^n \rho(i-\ell)_+ < 0 $, and the retirement benefit decreases accordingly; one can understand this as risk sharing to make the fund sustainable. The adjustment parameter $\theta$ determines the strength of the effects of this term, as  we have $\lim_{\theta \to +0} (1 - e^{-\theta}) = 0$ and  $\lim_{\theta \to \infty} (1 - e^{-\theta}) = 1$.    \\
	
	

	\item The \textcolor{black}{term (III)} shows the effects of the volatility of the fund's investment in the last $n$ years before the retirement; recall the definition of $\tsigma = \pi \sigma$ in \eqref{eq:expected-ann-log-return}. The adjustment parameter $\theta$ controls the influence of this volatility, as we have $\lim_{\theta \to +0}  {\rm (III)}  = 0$ and $\lim_{\theta \to \infty} {\rm (III)} = \tsigma \int_{i-n}^i \d Z(s)$. \\
\end{itemize} 

From these observations, one can understand that the adjustment parameter $\theta$ determines how strongly the retirement benefit is linked to the fund's actual investment performance. 
For a larger $\theta$, the terms \textcolor{black}{(II) and (III)} become more significant, and the retirement benefit is more directly influenced by the fund's investment performance. 
For a smaller $\theta$, the terms \textcolor{black}{(II) and (III)} become less significant, and the retirement benefit is determined mainly by the deterministic growth term \textcolor{black}{(I)}. 
This asymptotic analysis supports the informal discussion in Section \ref{sec:disccussion-indexation-rate} on the mechanism of the indexation rate.


One may conclude that a smaller $\theta$ may be more beneficial for individual participants, because it makes the retirement benefits less volatile. 
However, as discussed in Section \ref{sec:effects-adjust-fund-ratio}, a smaller $\theta$ makes the fund's operation more volatile, and thus a larger $\theta$ is more desirable for the fund's sustainability. 
Therefore,  $\theta$ should  be neither too small nor too large. 
We will discuss how to select the adjustment parameter $\theta$ (and the investment strategy $\pi$) in Section \ref{sec:BO}. 


\subsection{Effects of Intergenerational Risk Sharing}
\label{sec:comparison-pure-DC}


Lastly, we discuss the effects of IRS, by comparing the pension benefits of the IRS-DC plan and the corresponding pure DC plan. 
Because our focus is to understand how IRS works, we assume here that the pure DC plan uses the same investment strategy $\pi$ as the IRS-DC plan.
(Note that, in our numerical analysis in Section \ref{sec:numerical-analysis}, we consider this setting as well as the setting where the pure DC plan uses the optimal investment strategy.)

Consider two hypothetical individuals from generation $i \in \mathbb{N}$, who retire in the year $i$. 
One individual participates in the IRS-DC plan, and receives the retirement benefit \eqref{eq:CDC-benefit-genN}.
The other participates in the pure DC plan using the investment strategy $\pi$, and receives the retirement benefit denoted by $A_i(i)$. It is easy to see that $A_i(i)$ is given by
\begin{align}
	A_i(i) &=  c \sum_{n = 1}^N 
	\exp\bigg( \underbrace{n \tilde{\mu}}_{\rm \textcolor{black}{(I')}} + \underbrace{ \tsigma \int_{i-n}^i \d Z(s) }_{\rm \textcolor{black}{(II')}} \bigg). \label{eq:IDC-benefit-genN} 
\end{align}
By comparing   \eqref{eq:CDC-benefit-genN} and \eqref{eq:IDC-benefit-genN}, we can make the following observations: \\
\begin{itemize}
	\item The \textcolor{black}{term (I)} in  \eqref{eq:CDC-benefit-genN} and the term \textcolor{black}{(I')} in \eqref{eq:IDC-benefit-genN} are the same. \\
	
	\item The \textcolor{black}{term (II)} in  \eqref{eq:CDC-benefit-genN}, which represents the effects of the fund's surplus or deficit, does not exist in \eqref{eq:IDC-benefit-genN}. This is reasonable, because there is no IRS in the pure DC plan.  \\
	
	\item The \textcolor{black}{term (III)} in \eqref{eq:CDC-benefit-genN}, which shows the influence of the volatility of the investment, corresponds to the \textcolor{black}{term (II')} in \eqref{eq:IDC-benefit-genN}. Indeed, the term (III) converges to the term (II')  as $\theta \to \infty$. 
 However, one can see that the  \textcolor{black}{term (III)} is smaller than the \textcolor{black}{term (II')} for any fixed value of $\theta$. The smaller volatility in \eqref{eq:CDC-benefit-genN} is the result of IRS, and is controlled by the adjustment parameter $\theta$. \\    
	
\end{itemize}
This comparison describes how IRS works in the IRS-DC plan: 
{\em IRS reduces the volatility of investment returns (\textcolor{black}{term (III)} in \eqref{eq:CDC-benefit-genN}),  by letting the individuals share the fund's surplus or deficit  (\textcolor{black}{term (II)} in \eqref{eq:CDC-benefit-genN}).}
This effect of IRS is particularly important for protecting individual participants when the market is turbulent.
Our numerical analysis in Section \ref{sec:numerical-analysis} shows that IRS is beneficial in this way.

\section{Optimizing the Investment Strategy and Adjustment Parameter }\label{sec:BO}

We describe how to optimize the parameters of the IRS-DC pension model, namely the investment strategy $\pi$ and the adjustment parameter $\theta$,  
so as to maximize the welfare of pension participants.
In Section \ref{sec:expected-utility}, we first introduce an expected utility maximization problem that involves the welfare of all the generations including those from the future. 
Since there is no analytical solution for this maximization problem, we next explain how to solve it numerically using {\em Bayesian optimization} in Section \ref{sec:Bayesian-opt-main}.
We then describe the setting of simulations in Section \ref{sec:setting-experiments}, which will be used later in our numerical analysis.

\subsection{Expected Utility Maximization Problem}

\label{sec:expected-utility}

We consider a hypothetical social planner (fund manager) who decides the investment strategy $\pi$ and the adjustment parameter $\theta$ for the welfare of all the generations. 
To define the utility of this social planner, let  $U_\gamma: (0, \infty) \to (-\infty, \infty)$ be the constant relative risk aversion (CRRA) utility function:
\begin{equation} \label{eq:CRRA}
	U_\gamma(x) := \frac{x^{1-\gamma}}{1-\gamma},\quad\gamma > 0, \quad ( \gamma \not= 1),
\end{equation}
where $\gamma > 0$ is the level of relative risk aversion. 
We then define the utility of the social planner as the sum of discounted utilities of the retirement benefits for all the generations:
\begin{equation} \label{eq:utility-social-planner}
	\sum_{i=1}^{\infty}\beta^i U_\gamma(B_i(i)),
\end{equation}
where $\beta > 0$ is a discounting factor and $B_i(i)$ is the retirement benefit of the $i$-th generation who retires at time $t = i$; see Figure \ref{fig:cash-flow}, \eqref{eq:374} \textcolor{black}{ and} \eqref{dynamic:benefit}.



Lastly, we define our expected utility maximization problem as
\begin{align}\label{problem}
	& \max\limits_{\pi,\theta}\mathbb{E}\left[\sum_{i=1}^{\infty}\beta^i U_\gamma(B_i(i))\right] \quad \mbox{subject to}\quad    0 \leq \pi \leq 1,\quad   \theta > 0,
\end{align}
where the expectation is with respect to the retirement benefits $B_i(i)$ for all generations $i \in \mathbb{N}$. 
Recall that $B_i(i)$ are path-dependent and depend on the investment strategy $\pi$ and the adjustment parameter $\theta$. 

We numerically solve the maximization problem \eqref{problem}, since neither the expected utility in \eqref{problem} nor the solution for $\pi$ and $\theta$ are available in closed form. We approximate the expected utility in \eqref{problem} by Monte Carlo simulations, and optimize $\pi$ and $\theta$ using {\em Bayesian optimization}, as explained next.

\subsection{Bayesian Optimization for Expected Utility Maximization}
\label{sec:Bayesian-opt-main}

We briefly explain here how we use Bayesian optimization (BO)  for solving the expected utility maximization problem \eqref{problem}. 
For details, see Appendix \ref{sec:tutorial-BO} and references therein. 
BO is a modern machine learning approach for globally optimizing a black-box objective function, and has been shown to be more efficient than traditional approaches such as grid search \citep{shahriari2016taking}. 
It has been widely used in applications where the objective function is computationally expensive to evaluate, such as the optimization of hyper parameters of a large-scale AI model \citep{snoek2012practical}.  
The current work is the first attempt to apply BO in optimizing a pension system. 


The objective function in  \eqref{problem} takes $\pi$ and $\theta$ as an input and outputs the expected utility: 
$$
(\pi, \theta) \mapsto f(\pi, \theta) := \mathbb{E}\left[\sum_{i=1}^{\infty}\beta^i U_\gamma(B_i(i))\right],
$$
where we note again that $B_i(i)$ depends on $\pi$ and $\theta$.
The key idea of BO is to ``learn'' the landscape of the objective function $f(\pi, \theta)$ while searching for $\pi$ and $\theta$ that maximizes the objective function. 
BO first evaluates the function values $f(\pi, \theta)$ for some initial candidates of $\pi$ and $\theta$, and obtains a rough estimate for the landscape of $f(\pi, \theta)$. 
In the next step, BO finds $\pi$ and $\theta$ such that  the function value $f(\pi,\theta)$ and its {\em uncertainty} are both high, so as to balance the so-called {\em exploitation-exploration trade-off}.
BO then evaluates $f(\pi, \theta)$ for these $\pi$ and $\theta$, and updates the estimate of the landscape of $f(\pi, \theta)$.
BO iterates this learning-optimization procedure. 
Estimates of the maximizers, $(\pi^*, \theta^*) = \arg \max f(\pi, \theta)$ are obtained after a sufficient number of iterations \citep{Bul11}.

The above procedure is called ``Bayesian'' because the learning of the objective function is done by a Bayesian nonparametric method \citep{Rasmussen2006}.
The Bayesian method is used because it can yield both an estimate of the landscape as well as its uncertainties, which are crucial for the exploitation-exploration trade-off and for gaining the optimization efficiency. 
For implementation, we use the R package {\em mlrMBO} \citep{bischl2017mlrmbo} in our numerical analysis.


\subsection{Simulation Setting}
\label{sec:setting-experiments}

We explain here how we approximate the expected utility in \eqref{problem} by Monte Carlo simulations, which is necessary for applying Bayesian optimization. 
Moreover, we describe the problem setting for our numerical analysis in the next section.
First of all, we set the number of working generations as $N = 40$,  the discounting factor in \eqref{problem} as $\beta = 0.98$, and the annual contribution as $c = 1$. 

\subsubsection{Financial Market}

We consider three settings for the financial market that represent different market risks, to investigate when IRS  in the IRS-DC model works most effectively.
The market price of risk, a.k.a the Sharpe ratio, is defined by
\begin{equation}\label{sharperatio}
	\lambda:=\frac{\mu-r}{\sigma}
\end{equation}
where $\mu$ and $\sigma$ are the rate and volatility of the stock, and $r$ is the rate of the risk-free asset; see Section \ref{sec:financial-market}.
%
\textcolor{black}{
The Sharpe ratio quantifies the performance of a risky project in relation to a risk-free investment. It is one of the most frequently used performance measures, and we use it to describe different financial markets in our experiments.
Typically, a Sharpe ratio above 0.5 in the long run indicates great investment performance and is difficult to achieve, while a Sharpe ratio between $0.1$ and $0.3$ is often considered reasonable and can be achieved more easily \cite[e.g.,][]{sharpe1998sharpe}.
Table \ref{tab:my_sharpdata} shows the Sharpe ratios for different financial markets estimated from historical data.\footnote{\textcolor{black}{Since we only have limited access to the data of the Chinese market before 2003, we only report the Sharpe ratios for periods after 2003 for the Chinese market.  We report the Sharpe ratios for longer and shorter periods, the latter being a period around the financial crisis, to show that the Sharpe ratio may depend on the period considered.}} It shows that high values of the Sharpe ratio are around 0.3, and low values can be below $0.05$. }


\begin{table}
    \centering
    \captionsetup{font=normal}
    \captionsetup{width=.8\textwidth}
    \caption{\textcolor{black}{
        Sharpe ratios of different financial markets estimated by empirical data for three financial markets from the U.S., Germany and China. We use the stock market index and the 10-year treasury bond as surrogates for risky and risk-free assets. \textcolor{black}{We only have access to the data for the Chinese market after 2003}. For each market, the Sharpe ratio is reported for two periods: from 1992 (or 2003 for China) to 2021, and from 2003 to 2010, the latter being a period around the financial crisis. 
    }}
    \label{tab:my_sharpdata}
    \begin{tabular}{l|lll}
    \hline
    \hline
         Country:& U.S. &Germany  &China  \\
         \hline
         Stock:  & S\&P 500 & DAX & SSECI \\
         Bond: & 10y T-bond &10y T-bond & 10y T-bond\\
         \hline
         Period:  & 1992-2021& 1992-2021 &2003-2021\\
          Sharpe ratio:  &0.2827&0.2114 & 0.0631\\
         \hline
         Period: &2003-2010 &2003-2010 &  2003-2010\\
         Sharpe ratio: & 0.0798 & 0.3122 &0.0485 \\
         \hline
         \hline
    \end{tabular}
    \end{table}

For the simulation, we consider the following three settings for the financial market, with different levels of the Sharpe ratio: 
\begin{align*}
	& \mbox{Market 1:}\quad \lambda_1=\frac{\mu_1-r_1}{\sigma_1}=\frac{0.065-0.02}{0.15}=0.3;   \\
	& \mbox{Market 2:}\quad \lambda_2=\frac{\mu_2-r_2}{\sigma_2}=\frac{0.065-0.01}{0.25}=0.22; \\
	& \mbox{Market 3:}\quad \lambda_3=\frac{\mu_3-r_3}{\sigma_3}=\frac{0.065-0.01}{0.5}=0.11.
\end{align*}
We refer to Markets 1, 2, and 3 as M1, M2, and M3 for brevity. 
We call M1, M2, and M3 the markets with {\em high}, {\em intermediate}, and {\em low} Sharpe ratios, respectively. 
\textcolor{black}{
The calibrated values from the real world for the longer period (Table \ref{tab:my_sharpdata}) justify the use of the chosen Sharpe ratios in our experiment.}



\subsubsection{Risk Aversion of the Social Planner}

The relative risk aversion $\gamma$ in the the CRRA utility function \eqref{eq:CRRA} represents the social planner's risk attitude:  the social planner becomes more risk-averse if $\gamma$ is larger. 
To study the impacts of $\gamma$ on the optimal investment strategy $\pi$ and adjustment parameter $\theta$, we consider three settings: $\gamma = 3, 5, 10$. 


\subsubsection{Entry Cohorts}

The generations with indicators $i = 1, \dots, 40$ are those who participate in the IRS-DC plan at time $t = 0$, and are called {\em entry cohorts}.
For $t < 0$,  i.e., before participating in the IRS-DC plan, we assume that the entry cohorts participate in a pure DC plan that applies the optimal \emph{life-cycle investment strategy}, following \cite{gollier2008intergenerational}.
Namely, the pure DC plan invests a large amount into the stock when the participant is young and gradually reduces the amount invested in the stock as the participant approaches retirement.
To be more precise, for a generation $i$ where $i=1,\dots, 40$, let $B_i(t)$ be the individual account of generation $i$ and $Y_i(t)$ be the net present value at time $t$ of all the future contributions of generation $i$; then \textcolor{black}{the optimal fraction $\pi_i^{\rm ind}(t)$ of generation $i$'s wealth to be invested in the stock is given by}
\begin{equation}\label{eq:optimal individual}
	\pi_i^{\rm ind}(t) :=\pi^c\frac{B_i(t)+Y_i(t)}{B_i(t)}.
\end{equation}
See \citet[Eq. 71]{merton1975optimum}.
Notice that $\pi^c$ is the so-called {\em Merton constant} defined as
\begin{equation} \label{eq:Merton-const}
	\pi^c:=\frac{\lambda}{\gamma\sigma},
\end{equation}
where $\lambda$ is the Sharpe ratio in \eqref{sharperatio}.  Note that $Y_i(t)$ can be calculated straightforwardly here, as the interest rate risk is excluded. 

\begin{figure}
	\begin{subfigure}[b]{0.55\textwidth}
		\includegraphics[width=0.8\linewidth]{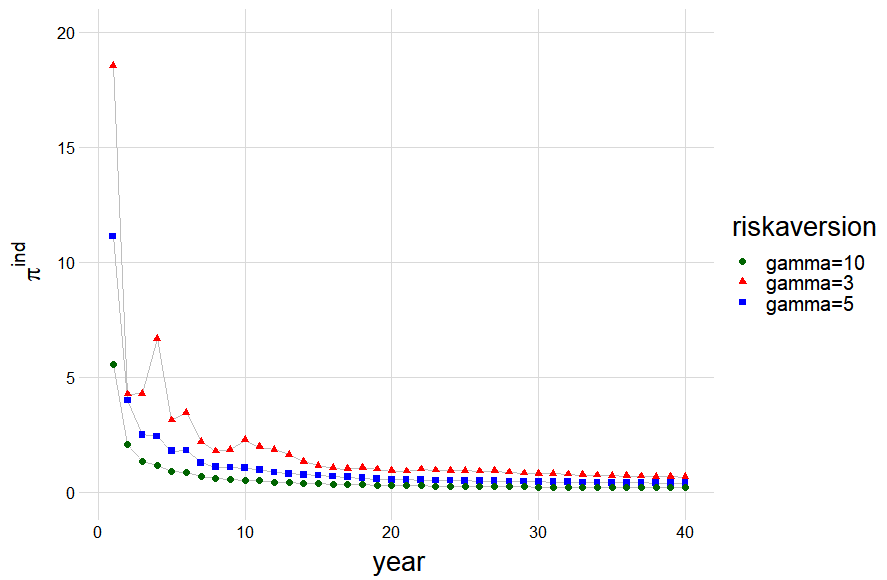}
		\vspace{0mm}
		\caption{Market 1}
	\end{subfigure}%
	\begin{subfigure}[b]{0.55\textwidth}
		\includegraphics[width=0.8\linewidth]{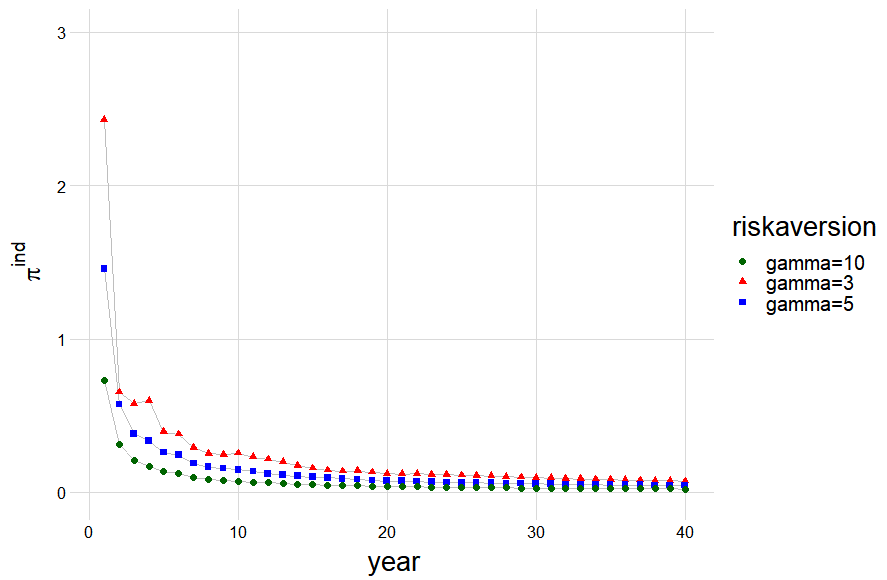}
		\vspace{0mm}
		\caption{Market 3}
	\end{subfigure}%
	\caption{Realizations of the optimal life-cycle investment strategy \eqref{eq:optimal individual} for an individual who participates in the pure DC plan for her entire working period of $40$ years. The left and right figures are for the Markets 1 and 3, respectively. Three values of the risk aversion $\gamma = 3, 5, 10$ are considered. The vertical axis represents \eqref{eq:optimal individual} and the horizontal axis the time.
	}\label{fig:life cycle}
\end{figure}

For an individual with the CRRA utility function, the life-cycle investment strategy \eqref{eq:optimal individual} provides the highest expected utility \citep{merton1975optimum,gollier2008intergenerational}. Hence, the life-cycle investment strategy and its variants have been popular choices for pure DC plans \citep[e.g.,][]{booth2000investment,haberman2002optimal}. 
%
%
%
Note that, when an individual is young, the discounted future income $Y_i(t)$ is much higher than her current wealth in the account $B_i(t)$, and thus $\pi^{\rm ind}(t)$ in \eqref{eq:optimal individual} is much larger than $1$. 
Therefore,  the life-cycle investment strategy \eqref{eq:optimal individual} implies a high-leverage (i.e., borrowing) strategy when the individual is young\footnote{\textcolor{black}{Note that \eqref{eq:optimal individual} is a random variable as $B_i(t)$ is a random variable. Therefore \eqref{eq:optimal individual} can increase in a short horizon of time, but in the long run, it decreases in expectation as $t$ increases, since as $t$ increases $Y_i(t)$ decreases and $B_i(t)$ increases in expectation. For more details on the life-cycle strategy, we refer to \citet{merton1975optimum}.}}    (see Figure \ref{fig:life cycle}).

The life-cycle investment strategy is also used in Section \ref{sec:experiment-CE} to make a comparison between the IRS-DC and the optimal pure DC plans.

\subsubsection{Euler-Maruyama Approximation} 
For simulating the dynamics of the asset process \eqref{eq:asset-345}, we use the Euler-Maruyama approximation. 
Given a finite time horizon $T > 0$, we divide the interval $[0, T]$ into $n$ equal time intervals:
$$[0,T]=[0,\Delta, 2\Delta,\cdots,n\Delta],
$$
where we set the step size as $\Delta = 1 / 12$, which corresponds to one month. 
Then we simulate the asset process as
\begin{align*}
	&A(t+\Delta) = A(t) + (\pi(\mu-r)+r)\Delta + \pi\sigma\sqrt{\Delta}Z, \nonumber\\
	&A(t)_+ =  A(t) + 40c - B_t(t),\quad \text{if} \; t  \in \mathbb{N} \cup \{0 \},
\end{align*}
where $Z$ is a standard normal random variable.  
The dynamics of the asset process in a pure DC plan is simulated in a similar way.  


The step size $\Delta = 1/12$ implies that the indexation rate \eqref{benefit-growth-rate} is adjusted monthly according to the funding ratio. This monthly update is more frequent than the annual cash flows of the fund. This setting reflects the fact that the market values of individual accounts usually vary more frequently than cash flows in reality. 
The IRS-DC fund is assumed to be fully funded at $t=0$ implying that the initial value of the notional funding ratio is one: $A(0)/L(0) = 1$.
(The influence of the initial funding ratio is examined in the numerical analysis in Section \ref{sec:funding-ratio-exp}.)


\subsubsection{Time Horizon $T$}

\label{sec:horizon-finite}

While the expected utility in  \eqref{problem} involves the infinite horizon, it is intractable for simulations and we need to use a finite horizon $T$. 
In our numerical analysis, we set the horizon as $T = 80$ years. 
For approximating the expected utility in \eqref{problem}, we then simulate 10,000 paths for the asset process until the horizon $T$ and compute the Monte Carlo average.
Note that, in this setting, the generations $i = 41, \dots, 80$ are those who spend their entire working periods in the IRS-DC plan.


\subsubsection{Upper Bound of the Adjustment Parameter $\theta$}

While the adjustment parameter $\theta$ can take an arbitrarily large value in theory, for numerical optimization of $\theta$ we need to set its upper bound. 
The range of $\theta$ is set as $0 < \theta \leq 1$ in our numerical analysis. 



\section{Numerical Analysis}
\label{sec:numerical-analysis} 

This section presents our numerical analysis  of the IRS-DC pension model.  
In Section \ref{sec:opt-invest-adjust}, we first discuss the optimal investment strategy and adjustment parameter obtained by Bayesian optimization.
In Section \ref{sec:funding-ratio-exp}, we then study the dynamics of the funding ratio, and discuss how the adjustment parameter $\theta$ affects its stability.
The stability of the funding ratio can be understood as the stability of the pension fund's operation. 
In Sections \ref{sec:ind-account-exp}, \ref{sec:distribution-benefits}, and \ref{sec:experiment-CE}, we focus on the individual accounts in the IRS-DC fund.  
We first study the dynamics of individual accounts in Section \ref{sec:ind-account-exp}, and then the distribution of retirement benefits in Section \ref{sec:distribution-benefits}. 
Lastly, we study the welfare of pension participants in Section \ref{sec:experiment-CE}.
\textcolor{black}{
Additional numerical analyses on a time-varying investment strategy and the influence of population structure are reported in Appendix \ref{sec:sec6}.
}


\subsection{Optimal Investment Strategy and Adjustment Parameter}

\label{sec:opt-invest-adjust}

\begin{table}
\centering
\captionsetup{font=normal}
\captionsetup{width=.8\textwidth}
	\caption{Optimal investment strategy $\pi^*$ and adjustment parameter $\theta^*$ obtained by Bayesian optimization for the 9 different settings, resulting from the 3 different values of the Sharpe ratio $\lambda$ and the 3 different values of the risk aversion $\gamma$. For comparison, the Merton constant is shown for each setting.
	}\label{table:opdesign}
	\begin{tabular}{cclllllllll}	
		\hline
		\hline
		&$\lambda=0.3$&&$\gamma=3$&&$\gamma=5$&&$\gamma=10$&\\
		\hline
		&$\pi^{*}:$&&$0.832$&&$0.519$&&$0.267$&\\
		
		&$\theta^{*}:$&&$1$&&$1$&&$1$&\\ 
		&{Merton constant}:&&${0.667}$&&${0.4}$&&${0.2}$&\\
		\hline
		&$\lambda=0.22$:&&$\gamma=3$&&$\gamma=5$&&$\gamma=10$\\
		\hline	&$\pi^{*}:$&&$0.479$&&$0.334$&&$0.124$\\
		&	$\theta^{*}:$&&$0.0651$&&$0.0535  $&&$0.0237$\\
		&	{Merton constant}:&&${0.293}$&&${0.176}$&&${0.088}$\\
		\hline
		&$\lambda=0.11$:&&$\gamma=3$&&$\gamma=5$&&$\gamma=10$\\
		\hline
		&$\pi^{*}:$&&$0.131$&&$0.06$&&$0.0544$\\
		&	$\theta^{*}:$&&$0.0835$&&$0.072 $&&$0.0000493$\\
		&	{Merton constant}:&&${0.073}$&&${0.044}$&&${0.022}$\\
		\hline
		\hline
	\end{tabular}
\vspace{5mm}
\end{table}

As explained in Section \ref{sec:setting-experiments}, we consider 9 different settings for the numerical analysis, resulting from 3 different values for the relative risk aversion ($\gamma=3, 5, 10$) of the social planner and 3 different values for the Sharpe ratio  ($\lambda=0.3, 0.22, 0.11$) of the financial market. 
In each setting, we find the optimal investment strategy $\pi$ and the adjustment parameter $\theta$ by Bayesian optimization, as described in Section \ref{sec:BO}.
We report the resulting optimal values of $\pi$ and $\theta$ in Table \ref{table:opdesign}. 
For comparison, we also report the Merton constant \eqref{eq:Merton-const} for each setting  in Table \ref{table:opdesign}, which will be used in the experiment in Section \ref{sec:experiment-CE}.

For each value of the Sharpe ratio $\lambda$,  the optimal $\pi^*$ and $\theta^*$ tend to decrease as the risk aversion $\gamma$ increases (with the exception of the case $\lambda = 0.3$, where $\theta^*$ remains $1$). 
Regarding the optimal investment strategy $\pi^*$, this tendency can be anticipated by the same tendency in the Merton constant \eqref{eq:Merton-const}, which is inversely proportional to the risk aversion $\gamma$.
Regarding the optimal adjustment parameter $\theta^*$, this tendency can be expected from the analysis in
Section \ref{sec:effects-adjust-pension-benefits}, where it is shown that a smaller adjustment parameter $\theta$ lowers the volatility of the retirement benefits; thus a higher risk aversion $\gamma$ leads to smaller $\theta^*$.

For each value of the risk aversion $\gamma$,  the optimal $\pi^*$ and $\theta^*$ tend to be smaller as the Sharpe ratio $\lambda$ becomes smaller (for the cases $\gamma = 3, 5$ and $\lambda = 0.22, 0.11$,  the adjustment parameter $\theta^*$ is comparably small).
One can understand this tendency in a similar way as the discussion in the above paragraph, since the Sharpe ratio represents the market price of risk. 
Note that $\theta^*$ is extremely small for $\gamma = 10$ and $\lambda = 0.01$, which implies the IRS-DC plan becomes similar to a DB plan, as discussed in Section \ref{sec:effects-adjust-pension-benefits}; in this case $\pi^*$ is also very small, meaning that the asset is mainly invested in the risk-free asset.


\subsection{Funding-Ratio Process  }
\label{sec:funding-ratio-exp}
\textcolor{black}{
 We next study the dynamics of the funding ratio $A(t)/L(t)$, investigating the influences of the adjustment parameter $\theta$ and the initial funding ratio $A(0) / L(0)$. 
 }

\subsubsection{Influence of the Adjustment Parameter}
\begin{figure}
\captionsetup{font=normal}
\begin{subfigure}[b]{0.55\textwidth}
\centering
\includegraphics[width=.7\linewidth]{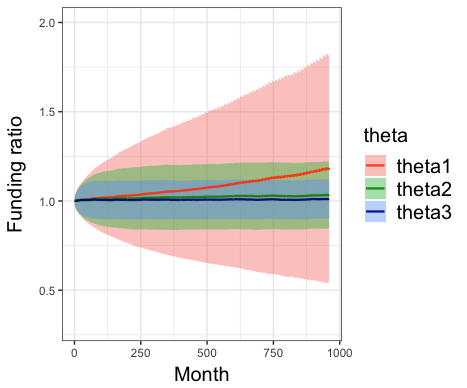}
\vspace{0mm}
\caption{
Means and standard deviations for $\theta_1 = 0.04$, $\theta_2 = 0.0835$ and $\theta_3 = 0.2$. 
}
\end{subfigure}%
\begin{subfigure}[b]{0.55\textwidth}
\centering
\includegraphics[width=.7\linewidth]{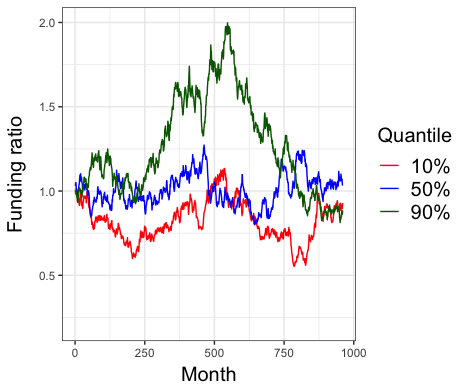}
\vspace{0mm}
\caption{
Paths for $\theta_1=0.04$.
}
\end{subfigure}%

\begin{subfigure}[b]{0.55\textwidth}
\centering
\includegraphics[width=.7\linewidth]{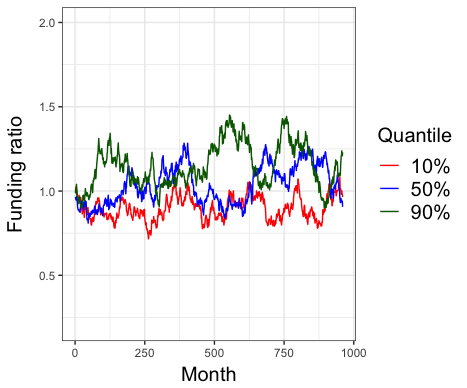}
\vspace{0mm}
\caption{
Paths for $\theta_2=0.0835$.
}
\end{subfigure}%
\begin{subfigure}[b]{0.55\textwidth}
\centering
\includegraphics[width=.7\linewidth]{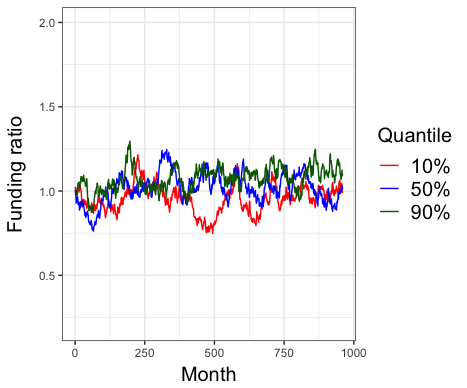}
\vspace{0mm}
\caption{
Paths for $\theta_3=0.2$.
}
\end{subfigure}%
\caption{
\textcolor{black}{
The influence of the adjustment parameter $\theta$ on the funding ratio process $A(t)/L(t)$ in Market 3. Panel (a) shows the mean and standard deviation of the funding ratio $A(t)/L(t)$ over 10,000 simulations as a function of time $t$, for $\pi = 0.131$ and each of three values of the adjustment parameter: $\theta_1 = 0.04$, $\theta_2 = 0.0835$ and $\theta_3 = 0.2$. Panel (b) shows the paths of $A(t)/L(t)$ with $\theta_1 = 0.04$  corresponding to the top 10\%, 50\% and 90\% values of the utility of the social planner over the 10,000 simulations. Panels (c) and (d) show those with $\theta_2 = 0.0835$ and $\theta_3  = 0.2$, respectively. 
}
}\label{fig:exprfun}
\end{figure}

\textcolor{black}{
We first examine the influence of the adjustment parameter $\theta$. We fix the investment strategy to $\pi = 0.131$, which is optimal for Market 3 with $\gamma = 3$ (see Table \ref{table:opdesign}). We consider three values for the adjustment parameter: $\theta_1 = 0.04$, $\theta_2 = 0.0835$ and $\theta_3 = 0.2$, where $\theta_2$ is optimal for Market 3 with $\gamma = 3$.  For each value of the adjustment parameter, we simulate the IRS-DC fund 10,000 times in Market 3; the results are summarized in Figure \ref{fig:exprfun}. 
}

\textcolor{black}{
Figure \ref{fig:exprfun} (a) shows the mean and standard deviation of the funding ratio $A(t)/L(t)$ over the 10,000 simulations as a function of time $t$, for each of the three values of $\theta$. The standard deviation is the smallest for $\theta_3 = 0.2$ and the largest for $\theta_1 = 0.04$; therefore the larger the adjustment parameter, the smaller the standard deviation of the funding ratio. This observation validates our analysis in Section \ref{sec:analysis}, which indicates that a larger adjustment parameter $\theta$ makes the funding ratio lower. Moreover, the mean of the funding ratio is close to $1$ for $\theta_2 = 0.0835$ and $\theta_2 = 0.2$, while the mean (and standard deviation) gradually increase for $\theta_3 =0.04$. This observation is also consistent with the analysis in Section \ref{sec:analysis}, which implies that a smaller adjustment parameter lets the funding ratio $A(t)/L(t)$ converge to $1$ more slowly.  
}

\textcolor{black}{
Figures \ref{fig:exprfun} (b), (c) and (d) show the paths of the funding ratio for three representative scenarios defined as follows. We pick up the three scenarios from the $10,000$ simulations that correspond to the top 10\%, 50\%, and 90\% values of the utilities of the social planner (see \eqref{eq:utility-social-planner}), and plot the funding ratio processes in these scenarios; these three scenarios can be interpreted as representing ``good'', ``medium'' and ``bad'' realizations of the financial market. The discrepancy between the paths of the funding ratio in these scenarios reduces as $\theta$ increases. This implies that the funding ratio volatility decreases as $\theta$ increases, and is consistent with our analysis in Section \ref{sec:analysis}.
}

\subsubsection{Influence of the Initial Funding Ratio}

\begin{figure}
\captionsetup{font=normal}
\begin{subfigure}[b]{0.55\textwidth}
\centering
\includegraphics[width=.7\linewidth]{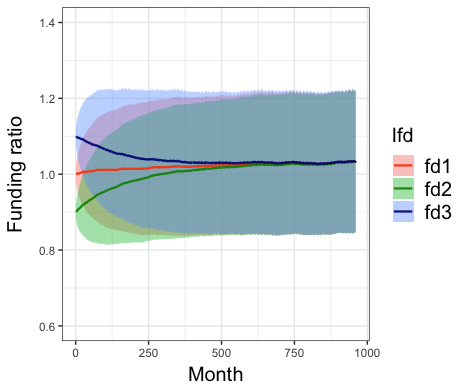}
\vspace{0mm}
\caption{}
\end{subfigure}%
\begin{subfigure}[b]{0.55\textwidth}
\centering
\includegraphics[width=.7\linewidth]{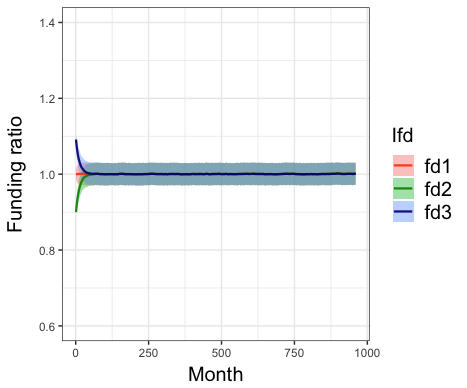}
\vspace{0mm}
\caption{}
\end{subfigure}%
\caption{
\textcolor{black}{
The influence of the initial funding ratio $A(0)/L(0)$ on the funding ratio process $A(t)/L(t)$. Panel (a) shows the mean and standard deviation of $A(t)/L(t)$ in Market 3 over 10,000 simulations, for $\pi = 0.131$ and each of three settings of the initial funding ratio: $A(0)/L(0) = 1$ (red), $A(0)/L(0) = 0.9$ (green; under-funded) and $A(0)/L(0) = 1.1$ (blue; over-funded). Panel (b) shows those in Market 1 with $\pi = 0.267$ and $\theta = 1$  and the same three values of the initial funding ratio.}
}\label{fig:fd_experiment}
\end{figure}

\textcolor{black}{
We next examine the influence of the initial funding ratio $A(0)/L(0)$ on the dynamics of the funding ratio $A(t)/L(t)$. We consider three cases for the initial funding ratio: (1) $A(0)/L(0) = 0.9$, (2) $A(0)/L(0) = 1$ and (3) $A(0)/L(0) = 1.1$. We simulate the IRS-DC fund 10,000 times for each case and calculate the mean and standard deviation of $A(t)/L(t)$. Figure \ref{fig:fd_experiment} shows the results for (a) Market 3 with $\pi = 0.131$ and $\theta = 0.0835$, which are optimal for $\gamma = 3$ in Market 3, and for (b) Market 1 with $\pi = 0.267$ and $\theta = 1$, optimal for $\gamma = 10$ in Market 1.  
Regardless of the initial funding ratio $A(0)/L(0)$, the mean of the funding ratio $A(t)/L(t)$ converges to 1 as $t$ increases. For Market 3, where the adjustment parameter $\theta$ is small, the mean of the funding ratio converges to $1$ slowly;  for Market 1, where the adjustment parameter is larger, the mean converges to $1$ immediately. Therefore these results suggest that the IRS-DC fund can self-stabilize the funding ratio to $1$, and a larger adjustment parameter $\theta$ leads to a quicker stabilization; again, this is consistent with the analysis in Section \ref{sec:analysis}. 
}


\subsection{Dynamics of Individual Accounts}
\label{sec:ind-account-exp}
\label{sec:stability-ind-account}

We next study the dynamics of individual accounts in the IRS-DC plan. 
As for the analysis in Section \ref{sec:comparison-pure-DC}, to study the effects of IRS, each individual account in the IRS-DC plan is compared with the corresponding account in a pure DC plan that uses the same investment strategy $\pi^*$ in Table \ref{table:opdesign}. (A pure DC plan using the optimal life-cycle investment strategy is compared in Section \ref{sec:experiment-CE}.) 
Since IRS is absent, the dynamics of an individual account in the pure DC plan is given by the asset process yielding \eqref{eq:IDC-benefit-genN}.



\begin{figure}
\captionsetup{font=normal}
\centering
\begin{subfigure}[b]{0.55\textwidth}
\includegraphics[width=.7\linewidth]{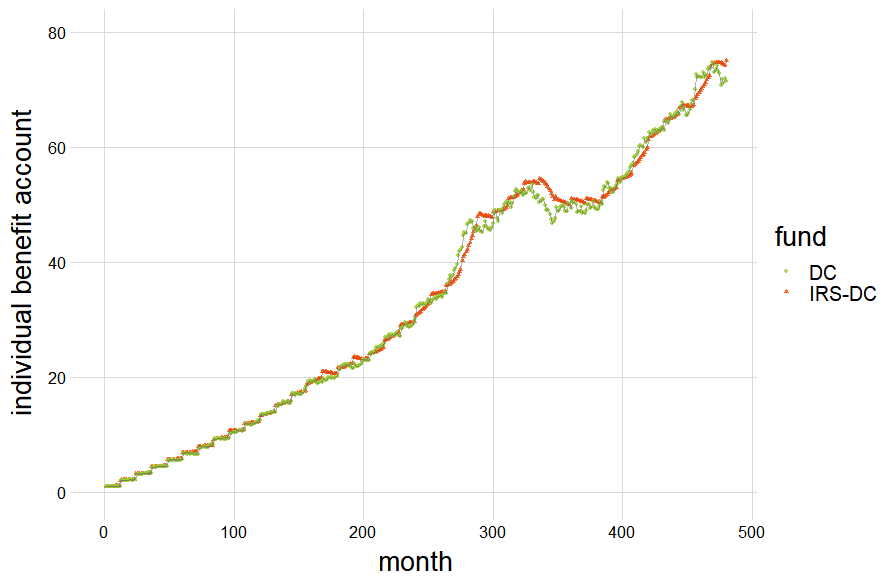}
\vspace{0mm}
\caption{Market 1.}
\end{subfigure}%
\begin{subfigure}[b]{0.55\textwidth}
\includegraphics[width=.7\linewidth]{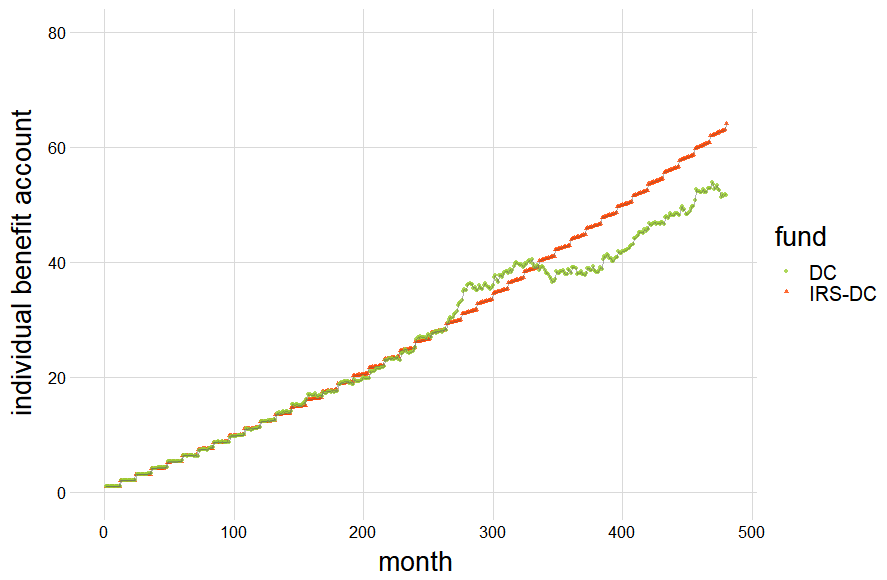}
\vspace{0mm}
\caption{Market 2}
\end{subfigure}%

\begin{subfigure}[b]{0.55\textwidth}
\includegraphics[width=.7\linewidth]{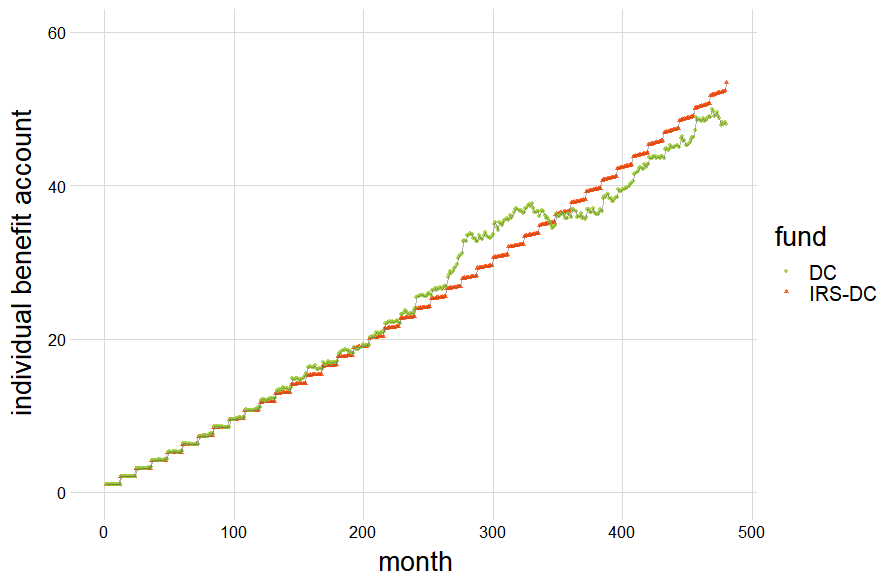}
\vspace{0mm}
\caption{Market 3.}
\end{subfigure}%
\caption{Examples of paths of the IRS-DC and pure DC accounts for the generation $i =41$ in the three market settings with risk aversion $\gamma = 10$. }
\label{fig:path}
\end{figure}

Figure \ref{fig:path} shows arbitrarily chosen paths of the individual accounts from the generation $i = 41$  in the IRS-DC and pure DC plans, for the three market settings and risk aversion $\gamma = 10$.
For Market 1, for which the adjustment parameter $\theta^*$ is large (see Table \ref{table:opdesign}), the paths of the IRS-DC and DC accounts are similar.
In contrast, for Markets 2 and 3, for which the adjustment parameter $\theta^*$ is smaller,  the IRS-DC account accumulates more stably than the pure DC account. 
This observation is consistent with the discussions in Sections \ref{sec:disccussion-indexation-rate}, \ref{sec:effects-adjust-pension-benefits} and \ref{sec:comparison-pure-DC}, where it is argued that a smaller adjustment parameter $\theta$ reduces the volatility of an IRS-DC account as a result of IRS.


We next quantify the effects of IRS on stabilizing the accumulation of an IRS-DC account. 
To this end, we calculate the {\em increment-ratio-based roughness (IR roughness)} \citep{bardet2011measuring},  a measure of roughness/smoothness of a stochastic process, for each path of the IRS-DC and pure DC accounts. 
The IR roughness takes a value between $0$ and $1$, and a larger value indicates that the path is smoother; see Appendix \ref{sec:IR-roughness} for details.
Table \ref{rough} shows the average of the IR roughness over the 10,000 simulations for each of the IRS-DC and pure DC accounts from the generation $i= 41$.
For all the 9 settings considered, the IRS-DC account has a larger IR roughness than the pure DC account, which implies that the IRS-DC account is smoother.
Therefore, IRS makes the accumulation of the IRS-DC account more stable than the pure DC account (Recall that the only difference between the IRS-DC and DC plans here is the existence of IRS).


\begin{table}
\begin{center}
\captionsetup{font=normal}
\caption{Average IR roughness over 10,000 simulations calculated for each of the IRS-DC and pure DC accounts, for the three market settings and the three levels of risk aversion.
}\label{rough}
\vspace{1mm}
\begin{tabular}{ccccc}
\hline
\hline
&IR roughness&$\gamma=3$&$\gamma=5$&$\gamma=10$\\
\hline	
Market 1 &IRS-DC&0.937&0.944&0.959\\
&DC&0.732&0.739&0.754\\

Market 2 &IRS-DC&0.993&0.996&1.000\\
&DC&0.731&0.735&0.752\\

Market 3 &IRS-DC&0.991&0.998&1.000\\
&DC&0.737&0.751&0.753\\
\hline
\hline
\end{tabular}
\end{center}
\end{table}

\subsection{Distribution of Retirement Benefits }

\label{sec:distribution-benefits}

\begin{figure}
\centering
\begin{subfigure}[b]{0.55\textwidth}
\centering
\includegraphics[width=.7\linewidth]{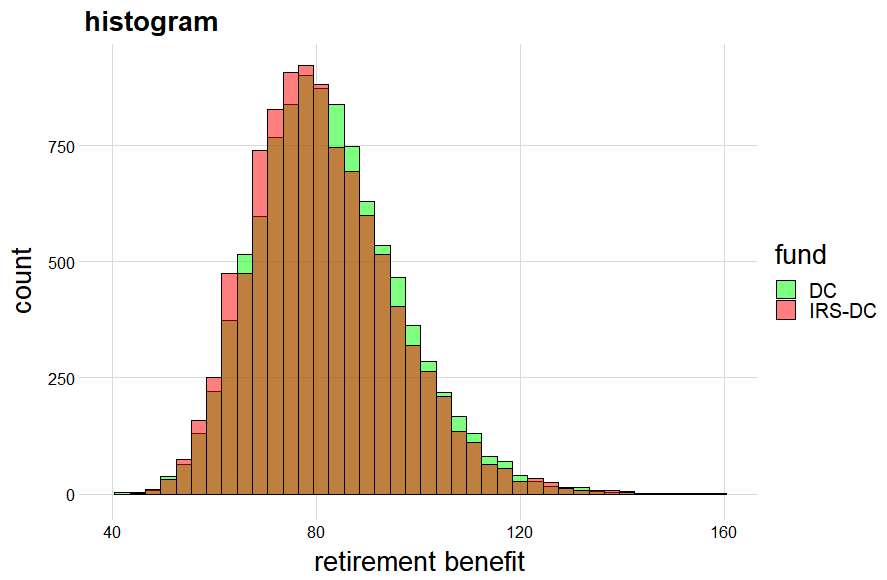}
\vspace{0mm}
\caption{Market 1}
\end{subfigure}%
\begin{subfigure}[b]{0.55\textwidth}
\centering
\includegraphics[width=.7\linewidth]{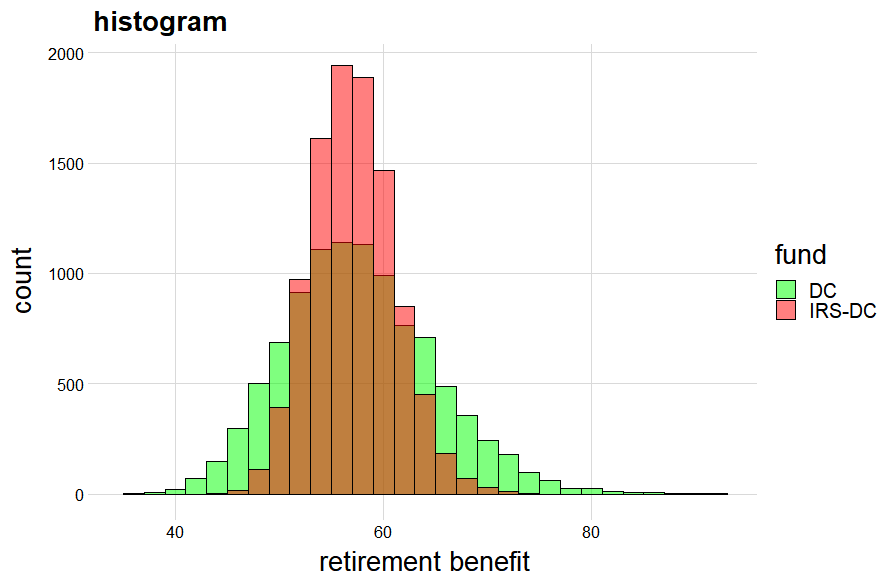}
\vspace{0mm}
\caption{Market 2}
\end{subfigure}%

\begin{subfigure}[b]{0.55\textwidth}
\centering
\includegraphics[width=.7\linewidth]{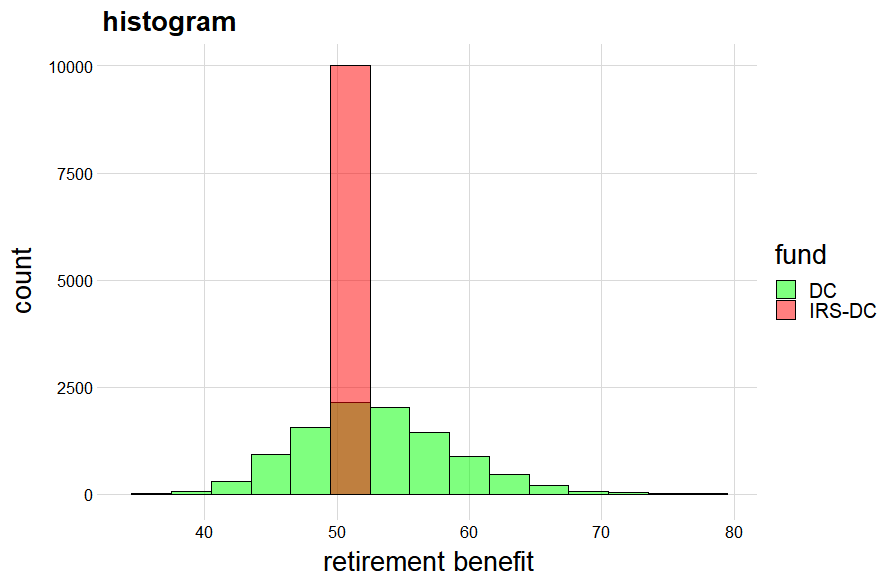}
\vspace{0mm}
\caption{Market 3}
\end{subfigure}%
\caption{Histograms of the retirement benefits (obtained from the 10,000 simulations) of the IRS-DC and pure DC accounts for the generation $i = 41$, for the three market settings and risk aversion $\gamma = 10$.  In each figure, the red and green histograms are those of the IRS-DC and pure DC benefits, respectively; the brown part shows the overlap between the two histograms. 
}
\label{fig:histogram}
\end{figure}

We next study how IRS affects the distribution of retirement benefits. 
Figure \ref{fig:histogram} shows the histograms of retirement benefits (from the 10,000 simulations) of the IRS-DC and pure DC accounts for the generation $i = 41$, for the three market settings and risk aversion $\gamma = 10$. (Results for $\gamma = 3, 5$ are shown in Appendix \ref{sec:appendix-dist-benefits}.) 
For Market 1, for which the adjustment parameter $\theta^*$ of the IRS-DC plan is large, the histograms of the IRS-DC and pure DC retirement benefits are almost identical.    
On the other hand, for Markets 2 and 3, for which the adjustment parameter is smaller, the volatility of the IRS-DC benefits is smaller than the pure DC benefits. 
In particular, for Market 3, for which the adjustment parameter is close to $0$, the volatility of the IRS-DC benefits is very small.   
These observations support the analysis in Sections \ref{sec:effects-adjust-pension-benefits} and \ref{sec:comparison-pure-DC} that a smaller adjustment parameter $\theta$ makes the retirement benefits less volatile by IRS.  
Moreover, our result is consistent with similar observations made by \citet{bams2016optimal} and \citet{donnelly2017discussion} that a collective DC scheme can reduce the volatility of retirement benefits.

\subsection{Welfare of Participants}
\label{sec:experiment-CE}

Lastly, we study how IRS can improve the welfare of the IRS-DC plan participants in terms of their expected utilities. 
To this end, we make a comparison with a pure DC plan that uses the \emph{optimal life-cycle investment strategy} in \eqref{eq:optimal individual}, which yields the highest expected utility for an individual investor.


For simplicity, we assume that each participant in the IRS-DC plan has the same CRRA utility $U_\gamma$ in \eqref{eq:CRRA} as the social planner.  
Similarly, to make a comparison straightforward, we assume that each participant in the pure DC plan has the same CRRA utility. 
To measure the welfare, we calculate the certainty equivalent (CE) for each participant.
That is,  for an IRS-DC participant from the generation $i$ with retirement benefit $B_i(i)$,  the CE is defined as the quantity $CE_{i}^{(\text{IRS-DC})} > 0$ satisfying
\begin{equation}\label{CE}
U_\gamma (CE_{i}^{(\text{IRS-DC})} ) = \mathbb{E}[U_\gamma (B_i(i))], 
\end{equation}
where we approximate the expectation in the right hand side by the empirical average of 10,000 realizations of $B_i(i)$. 
The CE of each participant in the pure DC plan is calculated similarly. 
Note that, since the utility function is strictly concave, the expected utility is monotonically increasing with respect to the CE; a higher CE implies a higher expected utility. 
We calculate the CEs of the participants in the IRS-DC plan and the pure DC plans  for the generations $i = 41, \dots, 80$.

\begin{figure}
\centering
\begin{subfigure}[b]{0.55\textwidth}
\centering
\includegraphics[width=0.8\linewidth]{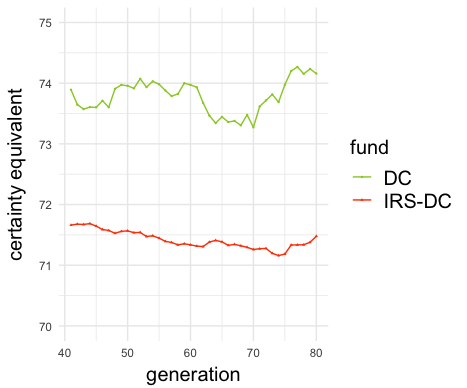}
\vspace{0mm}
\caption{Market 1.}
\end{subfigure}%
\begin{subfigure}[b]{0.55\textwidth}
\centering
\includegraphics[width=0.8\linewidth]{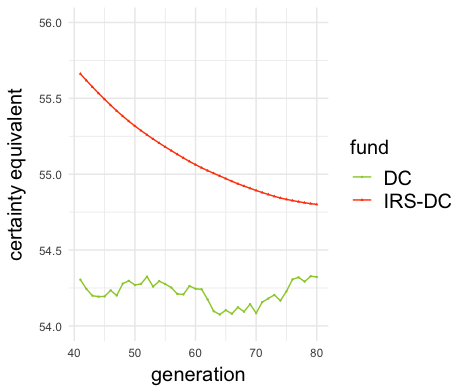}
\vspace{0mm}
\caption{Market 2.}
\end{subfigure}%

\vspace{5mm}
\begin{subfigure}[b]{0.55\textwidth}
\centering
\includegraphics[width=0.8\linewidth]{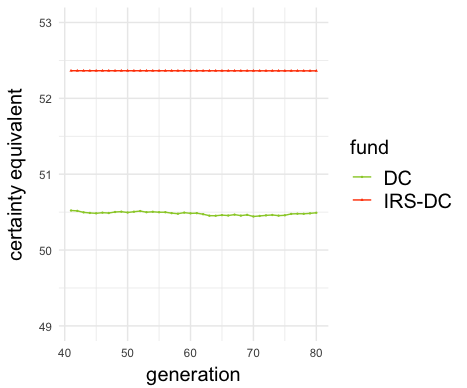}
\vspace{0mm}
\caption{Market 3.}
\end{subfigure}%
\caption{Certainty equivalents of the participants in the IRS-DC and pure DC plans for the generations $i = 41, \dots, 80$, for the three market settings and risk aversion $\gamma = 10$.
}\label{fig:opce}
\end{figure}

Figure \ref{fig:opce} shows the CEs of the IRS-DC and pure DC participants for the generations $i = 41, \dots, 80$, for the three market settings and risk aversion $\gamma = 10$. (Results for $\gamma = 3, 5$ are shown in Appendix \ref{sec:appendix-welfare}.)
For Market 1, where the Sharpe ratio is high, the pure DC participants obtain higher welfare than the IRS-DC participants.  
On the other hand, for Markets 2 and 3, where the Sharpe ratio is lower, the IRS-DC participants obtain higher welfare than the pure DC participants. 
This observation indicates that the IRS-DC plan can provide higher welfare than the optimal DC plan {\em when the market is more volatile} (in the sense of having a lower Share ratio).  
Therefore, IRS is expected to be particularly advantageous in protecting individual participants when the market is turbulent (e.g., when there is an economic shock).

While it has been generally known in the literature that IRS is welfare-improving,  there are a few key differences in our contribution.
To explain this, we make a comparison with closely related works.
\citet{gollier2008intergenerational} shows that IRS is welfare-improving over the optimal life-cycle investment strategy, but his analysis is based on the assumption that the pension fund can perform {\em borrowing} for investment (i.e., the investment strategy $\pi$ can be larger $1$); this assumption is not realistic for pension funds in reality. 
Moreover, his second-best strategy assumes the existence of a ``shareholder'' that helps finance the pension fund.   
Our result above shows that IRS can be welfare-improving even when borrowing is prohibited for the pension fund (i.e., $0 < \pi < 1 $) and without a shareholder.

\citet{cui2011intergenerational} show that a hybrid pension plan with IRS can provide higher welfare than a pure DC plan with an ``optimal'' investment strategy. 
However, their ``optimal''  individual investment strategy is not allowed to perform borrowing, and therefore it is less optimal than the optimal life-cycle strategy \eqref{eq:optimal individual}, which performs borrowing. Moreover, \citet{cui2011intergenerational} optimize the parameters of the pension fund so as to maximize the expected utility of {\em one specific entry cohort}, not all the generations; they then compare this entry cohort's welfare with a pure DC plan participant's welfare.  This way of optimizing the pension system is not appropriate as it ignores the other generations' welfare. On the other hand, we show that the IRS-DC plan, which optimizes for all the generations' utilities as in \eqref{problem}, can improve the welfare  over the optimal life-cycle investment strategy, when the market is volatile. 

\cite{bams2016optimal} consider a similar pension model as ours, but they do not show that their model can provide higher welfare than individual DC plans. Similarly, \citet{donnelly2017discussion} considers a related collective DC plan, but does not compare it with individual DC plans. \cite{chen2016intergenerational} study a three-pillar model in which the second pillar is a collective DC, DB, or hybrid pension plan, and make a comparison with the corresponding three-pillar model with the second pillar being an individual DC plan. While they show that the former yields higher welfare than the latter, they assume that the both plans use the {\em same} investment strategy, with the fraction invested in the stock being $\pi = 0.5$; therefore their individual DC plan is not optimal. 
Different from these previous works, we make a comparison with the optimal life-cycle investment strategy. 
By doing so, we show that the volatility of the financial market is a key factor that determines whether IRS is welfare-improving over the optimal life-cycle investment strategy.


\section{Concluding Remarks}\label{sec:conclusion}

We have shown that a fully funded collective DC pension system with intergenerational risk sharing (IRS) can improve the welfare of individual participants, as compared with individual DC benchmarks using the optimal life-cycle investment strategy, when the financial market is volatile.
Key new findings to the  literature include that i) the welfare improvement can be achieved without relying on borrowing and shareholders, in contrast to, e.g., \citet{gollier2008intergenerational}, and that ii) whether IRS improves the welfare depends on the volatility of the financial market, as measured by the Sharpe ratio.  These observations suggest that a fully funded pension system with a realistic investment strategy (i.e., without borrowing) can implement IRS and protect individual participants from a turbulent market.

Our investigation has been based on a stylized model, which we call the IRS-DC pension model, that uses an indexation rate of individual accounts as a device for IRS. This indexation rate, originally introduced by \citet{goecke2013pension}, is automatically adjusted according to the notional funding ratio of the pension fund, so as to balance the welfare of different generations and the sustainability of the pension fund. We have analyzed the funding ratio process and retirement benefits in the IRS-DC model, and how their volatility is controlled by the adjustment parameter in the indexation rate.  Moreover, we have shown how the adjustment parameter and the investment strategy can be optimized by using Bayesian optimization, a machine learning method for global optimization.



There are a number of possible future directions. 
First, as we have shown the effectiveness of the indexation rate of \citet{goecke2013pension} as a means for IRS in a collective pension system, the same indexation rate may be applied to other collective schemes, such as hybrid pension systems \citep[e.g.,][]{cui2011intergenerational,chen2016intergenerational} and notional DC pension systems \citep[e.g.,][]{settergren2001automatic}, where other forms of automatic adjustment rules are used for adjusting the individual accounts and/or contributions. This is worth investigating, as automatic adjustment rules have been used in real pension systems, such as the Dutch and Swedish pension systems \citep[Chapter 2]{paris2021pensions}.

Second, as Bayesian optimization provides an efficient way of optimizing the parameters of a pension system, it enables researchers to study optimal pension systems under more realistic but complex setups. For example, Bayesian optimization may be applied to optimize the three-pillar pension system of \cite{chen2016intergenerational}, which involves a number of parameters, by expected utility maximization; this may enable showing that their collective scheme is welfare-improving over the optimal individual benchmark using the life-cycle investment strategy, as we have shown for our collective scheme.

Third, our finding that IRS is welfare-improving in a volatile market is worth further investigation in a more realistic setup of the financial market.
While our setup of the Black-Scholes market (i.e., log-normally distributed stock returns)  follows many of related works  \citep[e.g.,][]{cui2011intergenerational,chen2016intergenerational}, it is known that this setup does not necessarily hold in reality \citep[e.g.,][]{cont2001empirical}. 
For example, the log returns of real stocks are known to have heavy tails, which implies that real financial markets are more volatile than the Black-Scholes market.  \textcolor{black}{Similarly,  it is more realistic to assume that the interest rate is stochastic and time-varying, rather than assuming a constant interest rate. 
Extending the current work to these more realistic settings will enable a deeper understanding of the functionality of the IRS.  }


Fourth, the discontinuity risk may be discussed for the IRS-DC model. 
We have implicitly assumed the mandatory participation of individuals by modelling that the population in each generation remains the same, as in related works \citep[e.g.,][]{gollier2008intergenerational,chen2016intergenerational}. One could relax this assumption by making the participation voluntary, and study individuals' preferences and how they impact the sustainability of the pension fund and the welfare of different generations \citep[e.g.,][]{beetsma2012voluntary}. 
Because contributions are not adjusted in the IRS-DC plan by design, it may be anticipated that the IRS-DC plan is less prone to discontinuity risk than DB-based pension plans. 
However, if voluntary participation changes the populations of different generations, the effectiveness of the IRS may be affected \textcolor{black}{(as suggested by the additional numerical analysis in Appendix  \ref{sec:sec62})}. It will be interesting to investigate whether mandatory participation is necessary for the IRS-DC plan to maintain effective IRS.


\section*{Acknowledgements}
We would like to express our gratitude to the editor and the anonymous reviewers for their time and insightful comments, which helped improve the paper. This work in part has been supported by the French government, through the 3IA Cote d’Azur Investment in the Future Project managed by the National Research Agency (ANR) with the reference number ANR-19-P3IA-0002, and by Deutsche Forschungsgemeinschaft with Grant number 418318744 of the research project: “Zielrente: die Lösung zur alternden Gesellschaft in Deutschland”.

\theendnotes
\newpage

\bibliographystyle{apalike}

\bibliography{zielrente}

\newpage

\appendix

\section{Proofs}

\subsection{Proof of Proposition \ref{lemma:funding-ratio} } \label{sec:proof-prop-fund-ratio}

\textcolor{black}{\begin{proof}
The identity \eqref{eq:lemma-cond-exp-290} follows from taking the conditional expectation of \eqref{eq:funding-ratio-331}, using that the Brownian motion $Z(s)$ for $t_0 < s < t$ is independent of the conditioning variable $\rho(t_0)_+$ and hence the conditional expectation of $Z(s)$ is zero.  
Eq.~\eqref{eq:lemma-cond-var-291} follows by using the Ito Isometry in \eqref{eq:funding-ratio-331}: 
\begin{align*}
    \mathbb{V}[\rho(t) \mid \rho(t_0)_+] & =   \tsigma^2  \mathbb{E} \left[   \left( \int_{t_0}^{t} e^{- \theta (t  - s)}  \d Z(s) \right)^2  \right]   =   \tsigma^2     \int_{t_0}^{t}   e^{- 2\theta (t  - s)}    \d s. 
\end{align*}
Eqs.~\eqref{eq:limit_0_condi_mean} and \eqref{eq:limit_0_condi_var} follow by taking the limits in \eqref{eq:lemma-cond-exp-290} and \eqref{eq:lemma-cond-var-291}.
\end{proof}}

\subsection{Proof of Proposition \ref{theo:benefits}} 

\label{sec:proof-theo-benefits}

\begin{proof}
Let $t_0 \in \mathbb{N}$ be such that $i-N \leq t_0 \leq i-1$. 
Let $\rho(t) = \ln (A(t) / L(t))$ be the log notional funding ratio.
By \eqref{eq:374}, \eqref{dynamic:benefit} and \eqref{benefit-growth-rate}, we have
\begin{align}
& B_i(t_0 + 1)  \\
&=  B_i(t_0)_+ \exp \left( \tilde{\mu} +  \theta \int_{t_0}^{t_0+1}  \rho(s) \d s\right) =  (B_i(t_0) + c) \exp \left( \tilde{\mu} +  \theta \int_{t_0}^{t_0+1}  \rho(s) \d s\right) , \nonumber \\
& \stackrel{(a)}{=}  (B_i(t_0) + c)  \exp \left( \tmu  + \rho(t_0)_+ - \rho (t_0+1) + \int_{t_0}^{t_0 + 1} \tsigma \d Z(s) \right), \nonumber \\
& \stackrel{(b)}{=}  (B_i(t_0) + c) \exp \left( \tmu  + (1 - e^{-\theta})\rho(t_0)_+   -   \int_{t_0}^{t_0+1} e^{- \theta (t_0 + 1  - s)} \tsigma \d Z(s) + \int_{t_0}^{t_0+1} \tsigma \d Z(s) \right), \nonumber  \\
& =  (B_i(t_0) + c) \exp \left( \tmu  + (1 - e^{-\theta})\rho(t_0)_+   + \tsigma  \int_{t_0}^{t_0+1} \left( 1 - e^{-\theta ( t_0 + 1 - s )}  \right) \d Z(s)  \right), \label{eq:expansion-925}
\end{align}
where  $(a)$ follows from \eqref{eq:log-fund-ratio-276} and $(b)$ follows from \eqref{eq:funding-ratio-331}.

We show a proof by induction. 
Suppose that, for $m \in \mathbb{N}$ with $0 < m \leq  N-1$, we have
\begin{align}
B_i( i - m ) =  
c \sum_{n= m + 1}^N  \exp  \bigg( &    (n - m)  \tmu  +    (1 - e^{-\theta}) \sum_{\ell=m+1}^{n} \rho(i - \ell )_+ \nonumber \\
& +   \tsigma \sum_{\ell = m+1}^{n} \int_{i - \ell}^{i-\ell + 1}  \left( 1 - e^{-\theta (i-\ell+1 - s)} \right) \d Z(s)   \bigg) \label{eq:932}
\end{align}
Note that the identity \eqref{eq:932}  holds for $m = N - 1$, since we have by \eqref{eq:expansion-925} and  $B_i(i-N)  = 0$  
\begin{align*}
& B_i( i - N + 1 )  = c  \exp  \bigg(  \tmu  +    (1 - e^{-\theta}) \rho(i-N)_+     +   \tsigma  \int_{i -  N}^{i - N  + 1}  \left( 1 - e^{- \theta (i - N + 1 -s )} \right) \d Z(s)   \bigg) . \nonumber
\end{align*}
By using \eqref{eq:expansion-925} with $t_0 = i - m$ , the assumption \eqref{eq:932} implies that
\begin{align*}
&B_i(i-m+1) = \\
&   (B_i(i - m) + c) \exp \left( \tmu  + (1 - e^{-\theta})\rho(i-m)_+   + \tsigma  \int_{i-m}^{i-m+1} \left( 1 - e^{-\theta ( i-m + 1 - s )}  \right) \d Z(s)  \right) \\
& =  \bigg[ c \sum_{n= m + 1}^N  \exp  \bigg(   (n - m)  \tmu  +    (1 - e^{-\theta}) \sum_{\ell=m+1}^{n} \rho(i - \ell )_+   \\
& \qquad \qquad \qquad \quad +   \tsigma \sum_{\ell = m+1}^{n} \int_{i - \ell}^{i-\ell + 1}  \left( 1 - e^{-\theta (i-\ell+1 - s)} \right) \d Z(s)   \bigg) + c \bigg] \\
& \quad \times \exp \left( \tmu  + (1 - e^{-\theta})\rho(i-m)_+   + \tsigma  \int_{i-m}^{i-m+1} \left( 1 - e^{-\theta ( i-m + 1 - s )}  \right) \d Z(s)  \right) \\
& = c \sum_{n= m}^N  \exp  \bigg(   (n - m + 1)  \tmu  +    (1 - e^{-\theta}) \sum_{\ell=m}^{n} \rho(i - \ell )_+   \\
& \qquad \qquad \qquad +   \tsigma \sum_{\ell = m}^{n} \int_{i - \ell}^{i-\ell + 1}  \left( 1 - e^{-\theta (i-\ell+1 - s)} \right) \d Z(s)   \bigg),
\end{align*}
which is the same expression as \eqref{eq:932}  with $m$ being replaced by $m - 1$. 
Therefore,  by induction, \eqref{eq:932} holds with $m  = 0$, which is \eqref{eq:CDC-benefit-genN}.
This completes the proof.

\end{proof}

\section{Tutorial on Bayesian optimization}

\label{sec:tutorial-BO}
We provide here a short tutorial on  Bayesian optimization (BO). For further details and references, see e.g.~\citet{shahriari2016taking}.

Let $\Omega$ be a parameter set and $f: \Omega \to \mathbb{R}$ be the objective function to be maximized. 
In our problem, this parameter set is $\Omega = [0,1] \times [0,1]$ and each $x := (\pi, \theta) \in \Omega$ represents a pair of the investment strategy $\pi$ and adjustment parameter $\theta$.
We define the objective function as the certainty equivalent (CE) of the expected utility in \eqref{problem}  with input parameters $x = (\pi, \theta)$:
\begin{align}
&f(x) := f(\pi, \theta) := CE (\pi, \theta), \label{eq:objective} \\
&\text{where}\quad CE (\pi, \theta) \geq 0\ \text{ is such that }\ U_\gamma (CE (\pi, \theta))  = \mathbb{E}\left[\sum_{t=1}^{\infty}\beta^{t} U_\gamma(B_t(t))\right].   \nonumber 
\end{align}
Note that the expected utility is a function of $x = (\pi, \theta)$, as the payment $B_t(t)$ depends on $\pi$ and $\theta$. 
Since the CRRA utility function $U_\gamma$ is strictly monotonically increasing with respect to its argument, the maximizer  of the certainty equivalent is the same as the maximizer of the expected utility: 
$$
\arg\max_{(\pi, \theta) \in \Omega} CE(\pi, \theta) = \arg\max_{(\pi, \theta) \in \Omega} \mathbb{E}\left[\sum_{t=1}^{\infty}\beta^{t} U_\gamma(B(t))\right].
$$
Thus, the maximization of the expected utility can be equivalently formulated as the maximization of the objective function \eqref{eq:objective}.

In our study, the expected utility is approximated by the Monte Carlo average of 10,000 simulations of the asset process $A(t)$ (and thus the resulting $B_t(t)$) for $t = 1, \dots, T := 80$.
Therefore, each evaluation of $f(x)$ for a given $x = (\pi, \theta)$ involves 10,000 simulations over 80 years on monthly basis, which is computationally expensive. 
If $A(t) \leq 0$ happens at any $t > 0$ for any of 10,000 simulations of the financial market, we set the objective function value to its minimum: i.e., $f(x) = 0$.


\subsection{Procedure of Bayesian optimization.}\label{sec:BO-basic}
First, we generate initial design points $x_1, \dots, x_{ n_{\rm init} }$ for some $n_{\rm init} \in \mathbb{N}$, and evaluate the function values $f(x_1), \dots, f(x_{n_{\rm init}} )$ on these points.
One can generate these initial points randomly (e.g, uniform sampling on $\Omega$) or deterministically (e.g., grid points). In our study, we use the design given by Latin hypercube sampling \citep{mckay2000comparison} on $\Omega$  with $n_{\rm init} = 10$.

Below we use the notation $D_n := \{ (x_i, f(x_i)) \}_{i=1}^n \subset \Omega \times \mathbb{R}$ to write the collection of points $x_1,\dots,x_n$ and the resulting function values $f(x_1), \dots, f(x_n)$. 
$D_n$ can be understood as ``data'' or ``observations'' about $f$ after $n$-time evaluations of the function. 
We also denote by $\alpha(x; D_n)$ the {\em acquisition function}, whose concrete form will be introduced later in Section \ref{sec:acquisition}.
The acquisition function $\alpha(x; D_n)$ is a function of $x \in \Omega$ and defined from $D_n$. 

BO iterates the following procedure for $n = n_{\rm init} + 1, n_{\rm init} + 2, \dots, M$, \textcolor{black}{where $M$ is the total number of function evaluations.} 
\begin{enumerate}
\item Compute $x_{n + 1} \in \arg\max_{x \in \Omega} \alpha(x; D_n)$,
\item Simulate $f(x_{n+1})$, and augment the data $D_{n+1} := D_n \cup \{ (x_{n+1}, f(x_{n+1})) \}$.
\end{enumerate}
An estimate of the optimal parameters is then given as the maximizer from the evaluated inputs $x_1, \dots, x_M$: 
$$
x^* \in \arg \max \left\{ f(x) \mid x \in \{ x_1, \dots, x_M \} \right\}
$$

The acquisition function $\alpha(x; D_n)$ determines the next point $x_{n+1}$ to evaluate the objective function $f$. Note that the computational cost of solving $\max_{x \in \Omega}\alpha(x; D_n)$ is negligible compared to the computational cost of evaluating $f(x_{n+1})$, as $\alpha(x; D_n)$ can be evaluated cheaply. 

The acquisition function is designed so as to balance the {\em exploitation} and {\em exploration}. Exploitation is a strategy to search for in a region near the current maximizer in $x_n^* := \arg\max \{ f(x) \mid x \in \{x_1,\dots,x_n \} \}$; exploration is to search for in a region far from the evaluated points $x_1,\dots,x_n$. This {\em exploration-exploitation trade-off} is enabled by the learning and uncertainty quantification of the response surface of $f$ from the data $D_n$.
This is done by Gaussian process regression, which we will explain next.

\subsection{Gaussian process regression} 
\label{sec:GP-regression}

\begin{figure} 
\centering
\includegraphics[width=0.8\linewidth]{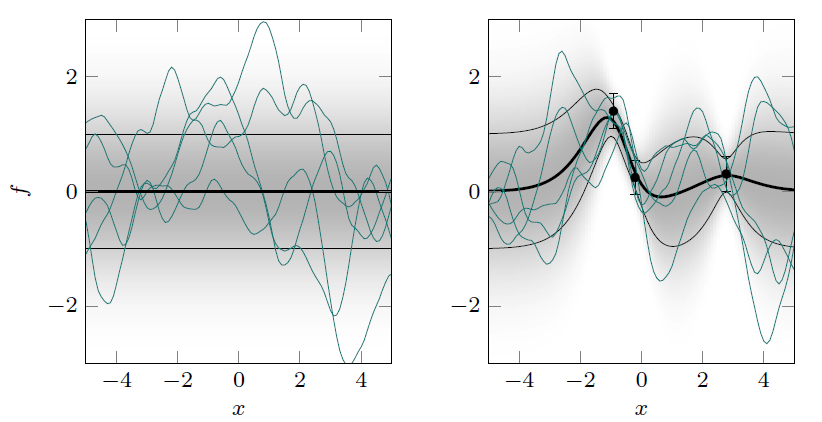}
\caption{Illustrations of Gaussian process regression, adapted from \citet[Figure 1]{kanagawa2018gaussian}.
\textbf{Left}: The green curves are 5 sample paths from the prior Gaussian process \eqref{eq:prior-GP} using the Mat\'ern kernel \eqref{eq:matern} with $h = 1$. The thick black line is the prior mean function $m(x)=0$. \textbf{Right}: the three points are noise-perturbed observations $(x_i, y_i)_{i=1}^3$, where $y_i = f^*(x_i) + \varepsilon_i$ with $f^*$ being the ground-truth function and $\varepsilon_i$ being an independent zero-mean Gaussian noise with variance $\sigma^2 := 0.01$.
The thick black curve is the posterior mean function $m_n(x)$ in \eqref{eq:GP-post-mean} and the two thin black curves are the posterior standard deviation function $\sigma_n(x) = \sqrt{k_n(x,x)}$ in \eqref{eq:gp-std}, in which ${\bf K}_{n}^{-1} $ is replaced by $({\bf K}_{n} + \sigma^2 {\bf I})^{-1}$ and ${\bf f}_n := (y_1, y_2 ,y_3)^\top$. (These modifications are theoretically justified; see e.g.~\citet{kanagawa2018gaussian} and references therein for details.) The green curves are 5 sample paths from the posterior Gaussian process \eqref{eq:GP-posterior}} 
\label{fig:gp-illustration}
\end{figure}

Gaussian process regression \citep{Rasmussen2006} is a Bayesian non-parametric method for learning (or approximating) an unknown function $f:\Omega \to \mathbb{R}$ from its finite observations (data) $D_n = \{ (x_i, f(x_i)) \}_{i=1}^n$. 
Recall that Bayesian inference in general proceeds as follows:
a) define a {\em prior distribution} for the quantity of interest, b) collect observations (data) related to that quantity, and c) update the prior distribution to the {\em posterior distribution} using the observed data, applying Bayes' rule. 
In Gaussian process regression, the quantity of interest is the unknown {\em function} $f$, and a') one defines a prior distribution of $f$ as a Gaussian process (or Gaussian random field), b') collects data $D_n = \{ (x_i, f(x_i)) \}_{i=1}^n$, and c') updates the prior Gaussian process to the {\em posterior} Gaussian process, applying Bayes' rule.
See Figure \ref{fig:gp-illustration} for illustrations of Gaussian process regression.

\subsubsection{Prior Gaussian Process}
A Gaussian process is completely specified by its {\em mean function}  $m: \Omega \to \mathbb{R}$ and {\em covariance function} $k: \Omega \times \Omega$.  
We write $f \sim \mathcal{GP}(m, k)$ to mean that $f$ is a sample path of the Gaussian process with mean function $m$ and covariance function $k$.
Then we have $m(x) = \mathbb{E}[f(x)]$, $x \in \Omega$ and $k(x,x') = \mathbb{E}[ (f(x) - m(x)) (f(x') - m(x')) ]$, $x, x' \in \Omega$. 
By specifying $m$ and $k$, we implicitly specify the corresponding Gaussian process. 

For simplicity, we consider a Gaussian process with the zero-mean function (i.e., $m(x) = 0$, $\forall x \in \Omega$) for our prior distribution of the objective function $f$:
\begin{equation} \label{eq:prior-GP}
f \sim \mathcal{GP}(0, k).
\end{equation}
What we need is to specify the covariance function $k$.
By doing so, we can express our assumption or knowledge regarding key properties of the objective function $f$, such as its smoothness and structure.

Popular choices of covariance kernels include square-exponential kernel $k(x,x') = \exp(- \| x - x' \|^2 / h)$ with $h > 0$ and {\em Mat\'ern kernels}. 
In our study we use the so-called Mat\'ern-$5/2$ kernel of the form
\begin{equation} \label{eq:matern}
k(x, x') = \left( 1 + \frac{\sqrt{5} \| x - x' \|}{h} + \frac{5 \| x - x' \|^2}{3 h^2} \right) \exp \left( - \frac{\sqrt{5} \| x - x' \| }{h} \right)
\end{equation}
where $h > 0$ is a scale parameter.\footnote{In our simulation study, we use the default value for $h$ of the mlrMBO package.}
Roughly, this kernel leads to $f \sim \mathcal{GP}(0, k)$ that is almost surely twice differentiable \cite[e.g.,][Section 4.4]{kanagawa2018gaussian}. Thus, with this kernel we essentially assume this degree of smoothness for the objective function, and this is our prior assumption.


\subsubsection{Posterior Gaussian Process}
The use of a Gaussian process as a prior leads to an {\em analytic} expression of the resulting posterior distribution.
Given data $D_n = \{ (x_i, f(x_i)) \}_{i=1}^n$, the posterior distribution of $f$ is also given as a Gaussian process
\begin{equation} \label{eq:GP-posterior}
f | D_n \sim \mathcal{GP} (m_n, k_n),
\end{equation}
where $m_n: \Omega \to \mathbb{R}$ is the {\em posterior mean function} and $k_n: \Omega \times \Omega \to \mathbb{R}$ is the {\em posterior covariance function}, given by
\begin{align}
m_n(x) & = \mathbb{E}[f(x) | D_n] =  {\bf f}_n^\top {\bf K}_n^{-1} {\bf k}_n(x), \quad x \in \Omega, \label{eq:GP-post-mean} \\
k_n(x,x') &= \mathbb{E}[ ( f(x) - m_n(x) ) ( f(x') - m_n(x) ) | D_n ] \nonumber \\
& = k(x,x') - {\bf k}_n(x)^\top {\bf K}_n^{-1} {\bf k}_n(x'), \quad x, x' \in \Omega,  \label{eq:GP-post-cov}
\end{align}
where ${\bf f}_n := (f(x_1), \dots, f(x_n))^\top$, ${\bf k}_n(x) := (k(x,x_1), \dots, k(x, x_n))^\top \in \mathbb{R}^n$ and ${\bf K}_n := (k(x_i, x_j))_{i, j = 1}^n \in \mathbb{R}^{n \times n}$.
For the detail of the above derivation, see \cite{Rasmussen2006}.

The posterior mean function $m_n$ in \eqref{eq:GP-post-mean} is an approximation of the objective function $f$ based on the data $D_n$. It works as a computationally cheaper surrogate model of $f$. On the other hand, the {\em posterior standard deviation} 
\begin{equation} \label{eq:gp-std}
\sigma_n(x) := \sqrt{k_n(x,x)} = \sqrt{\mathbb{E}[ (f(x) - m_n(x))^2 | D_n] } 
\end{equation}
quantifies the uncertainty about the unknown function value $f(x)$. 
These $m_n$ and $\sigma_n$ are the building blocks of the acquisition function, as we will see next.

\subsection{Acquisition function} \label{sec:acquisition}

We now introduce the concrete form acquisition function $\alpha(x; D_n)$.
There are many acquisition functions proposed in the literature; see \citet[Section IV]{shahriari2016taking}.
Most popular ones include the {\em EI} (Expected Improvement), {\em GP-UCB} (Gaussian Process Upper Confidence Bound), and {\em ES} (Entropy Search). 
In this paper, we use the EI acquisition function, which is standard and theoretically well studied \citep{Bul11}. 
Let 
$$
f_n^* := \max_{i = 1, \dots, n} f(x_i), \quad x_n^* \in \arg \max \{ f(x) \mid x \in \{x_1, \dots, x_n\} \}.
$$
be the maximum and the maximizer of the objective function $f(x)$ over the currently evaluated inputs $x_1, \dots, x_n$.
The EI acquisition function $\alpha(x;D_n)$ at $x$ is defined as the expected improvement of the function value $f(x)$ over the current maximum $f^*_n$, where the expectation is with respect to the posterior Gaussian process \eqref{eq:GP-posterior}:
\begin{align}
a(x; D_n) &:= \mathbb{E}_{ f \sim \mathcal{GP}(m_n, k_n)}  [ \max(f(x) - f^*_n, 0  )   ]  \nonumber \\
&=  \underbrace{ \sigma_n(x) \phi\left(\frac{m_n(x) - f^*_n}{\sigma_n(x)} \right) }_{\rm Exploration} + \underbrace{(m_n(x) - f^*_n) \Phi\left(\frac{m_n(x) - f^*_n}{\sigma_n(x)} \right)}_{\rm Exploitation}, \label{eq:EI_formula}
\end{align}  
where $\phi:\mathbb{R} \to [0, \infty)$ is the probability density function of a standard Gaussian random variable, and $\Phi: \mathbb{R} \to [0, 1]$ is its cumulative distribution function: $\Phi(y) := \int_{-\infty}^y \phi(s) ds$, $y \in \mathbb{R}$.

The first term in \eqref{eq:EI_formula} represents the {\em exploration}, as it becomes large when $\sigma_n(x)$, which represents the uncertainty about the function value $f(x)$, is large.  This is typically the case when $x$ is far from already evaluated locations $x_1,\dots,x_n$. 
The second in \eqref{eq:EI_formula} represents the {\em exploitation}, as it becomes large when $m_n(x) - f_n^*$ is large and $\sigma_n(x)$ is small. This is typically the case when $x$ is near the current maximizer $x_n^*$. Thus, the EI acquisition function naturally balances the exploration-exploitation trade-off, and the next point $x_{n+1} \in \arg \max_{x \in \Omega} \alpha(x; D_n)$ achieves such a balance.


\subsection{Demonstration} \label{sec:BO-demo}

\begin{figure}
\centering
\includegraphics[width=0.45\linewidth]{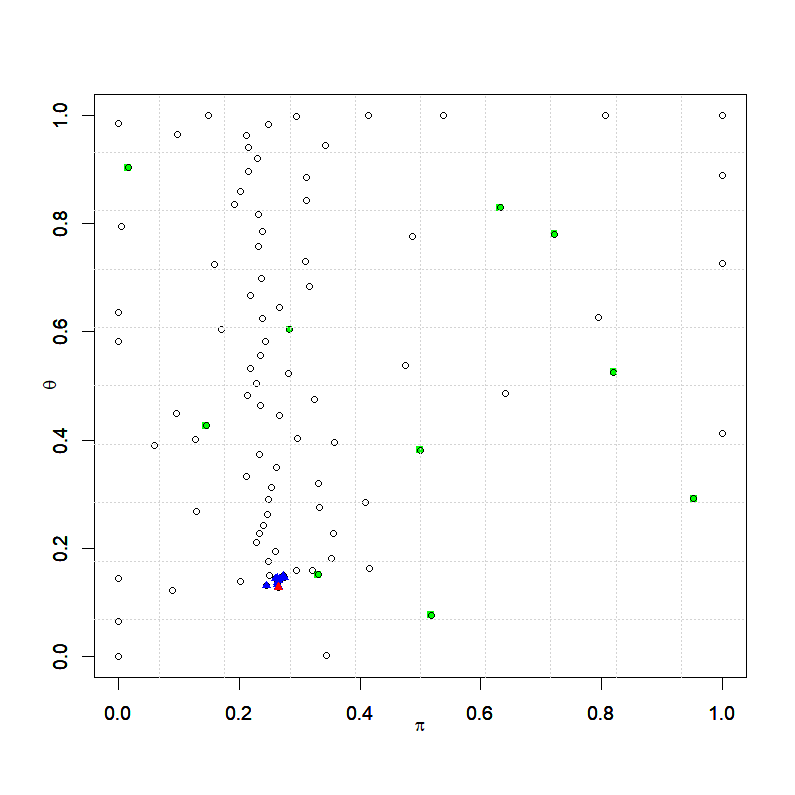}
\caption{\emph{Demonstration of Bayesian optimization with $\gamma = 10$ (see Section \ref{sec:BO-demo}).The points represent the values of $(\pi, \theta)$ evaluated by Bayesian optimization. The green points are the initial 10 points given by Latin hypercube sampling. The red point is the maximizer found by Bayesian optimization after 100 evaluations, and the blue points (largely overlapping the red point) are the 10 second-best points. }}
\label{fig:BO-demo-gamma10}
\end{figure}

Figure \ref{fig:BO-demo-gamma10} shows an example of points $x = (\pi, \theta)$ evaluated by BO for $\gamma = 10$. The green points are the $n_{\rm init} = 10$ initial design points $x_1, \dots, x_{n_{\rm init}}$ generated by Latin hypercube sampling. The total number of evaluated points is $M = 100$. The red point is the maximizer $x_M^*$, and the blue points are the 10 other second-best parameters (largely overlapping with the red point). For a comparison, we show $10 \times 10$ grid points.

\section{IR-roughness Measure}
\label{sec:IR-roughness}

To describe the IR-roughness measure \citep{bardet2011measuring}, we suppose that the path of each individual account is represented by a function $h : [0, \widetilde{T}] \to \mathbb{R}$, where $\widetilde{T}=40$ is its terminal time. Note that $\Tilde{T}$ is the terminal time for one generation but is different from the terminal time for the operation of the pension fund $T$. Discretizing the domain to $n - 1  \in \mathbb{N}$ intervals, the first-order IR-roughness is defined as
\begin{equation}\label{roughness}
R^{1,n}(h):=\frac{1}{n-1}\sum_{j=0}^{n-2}\frac{\left|h( {\color{black}{\widetilde{T}} }\frac{j+1}{n})-h(\widetilde{T}\frac{j}{n})+h(\widetilde{T}\frac{j+2}{n})-h(\widetilde{T}\frac{j+1}{n})\right|}{\left|h(\widetilde{T}\frac{j+1}{n})-h(\widetilde{T}\frac{j}{n})\right|+\left|h(\widetilde{T}\frac{j+2}{n})-h(\widetilde{T}\frac{j+1}{n})\right|}.
\end{equation} 
By the triangle inequality, the numerator in the sum is less than or equal to the denominator, and thus $R^{1,n}(h)$ takes values between $0$ and $1$. When the signs of the two increments $h( \widetilde{T}(j+1)/n ) - h( \widetilde{T} j/n ) $ and  $h( \widetilde{T}(j+2)/n ) - h( \widetilde{T}(j+1)/n )$ are the same, the numerator equals the denominator; when those signs are different, the numerator is smaller than the denominator. As such, $R^{1,n}(h)$ reflects the sign changes of the function $h$ and thus quantifies its roughness. Intuitively, $R^{1,n}(h)$ is close to $0$ when $h$ is rough, and is close to $1$ when $h$ is smooth.  In fact, \cite{bardet2011measuring} show that, for a sufficiently smooth $h$, $R^{1,n}(h)$ converges to $1$ as $n \to \infty$.

\section{Supplementary Numerical Results}

We show here additional numerical results not included in the main body of the paper.

\subsection{Distribution of Retirement Benefits}  
\label{sec:appendix-dist-benefits}

\begin{figure} \centering
\begin{subfigure}[b]{0.55\textwidth}
\includegraphics[width=0.9\linewidth]{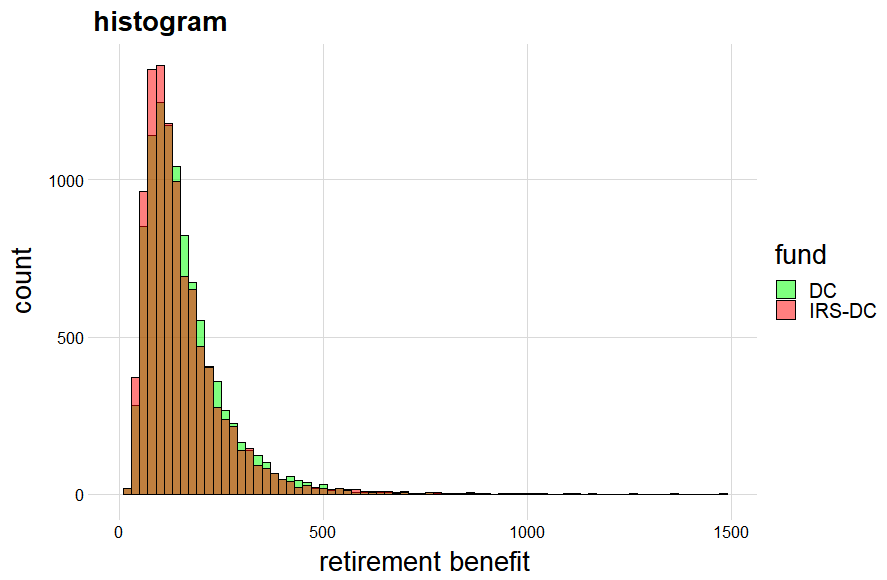}
\vspace{0mm}
\caption{Market 1: $\gamma=3$}
\end{subfigure}%
\begin{subfigure}[b]{0.55\textwidth}
\includegraphics[width=0.9\linewidth]{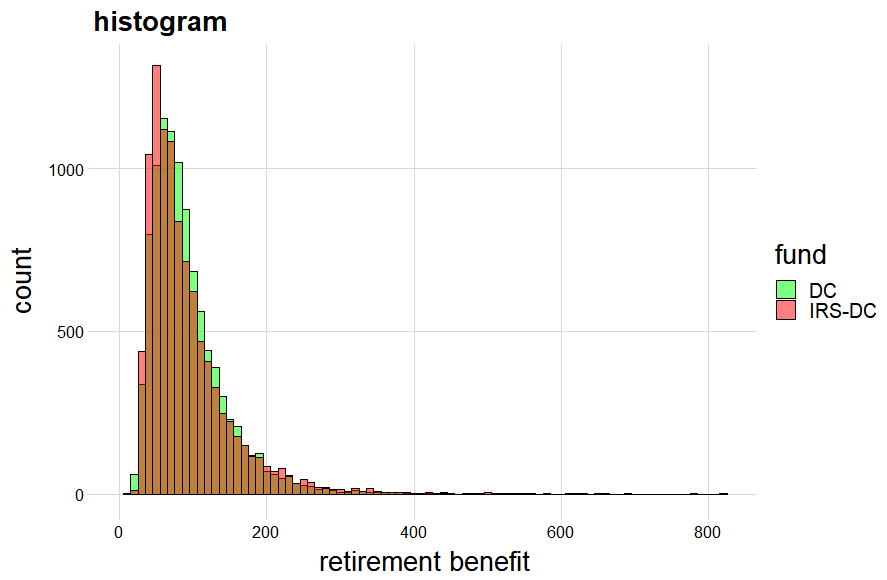}
\vspace{0mm}
\caption{Market 2: $\gamma=3$}
\end{subfigure}%

\vspace{5mm}
\begin{subfigure}[b]{0.55\textwidth}
\includegraphics[width=0.9\linewidth]{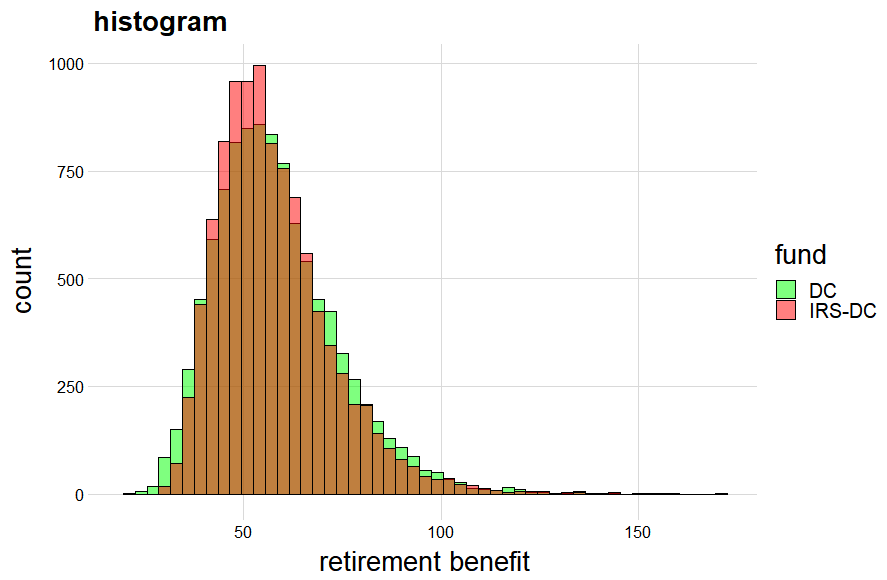}
\vspace{0mm}
\caption{Market 3: $\gamma=3$}
\end{subfigure}%
\begin{subfigure}[b]{0.55\textwidth}
\includegraphics[width=0.9\linewidth]{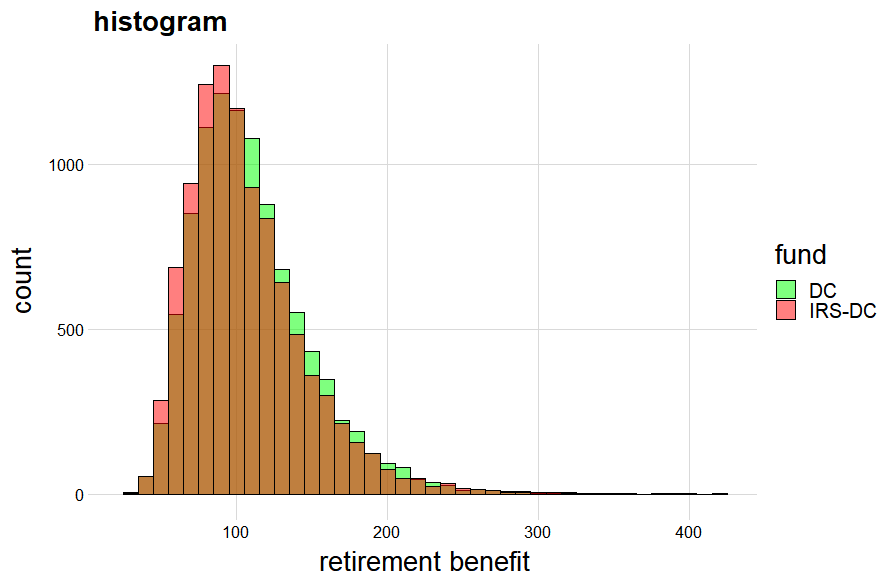}
\vspace{0mm}
\caption{Market 1: $\gamma=5$}
\end{subfigure}%

\vspace{5mm}
\begin{subfigure}[b]{0.55\textwidth}
\includegraphics[width=0.9\linewidth]{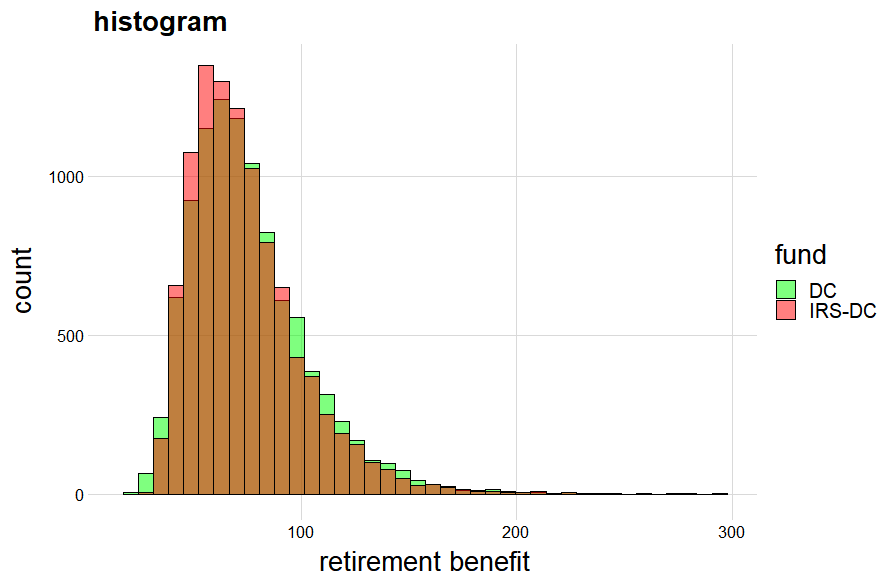}
\vspace{0mm}
\caption{Market 2: $\gamma=5$}
\end{subfigure}%
\begin{subfigure}[b]{0.55\textwidth}
\includegraphics[width=0.9\linewidth]{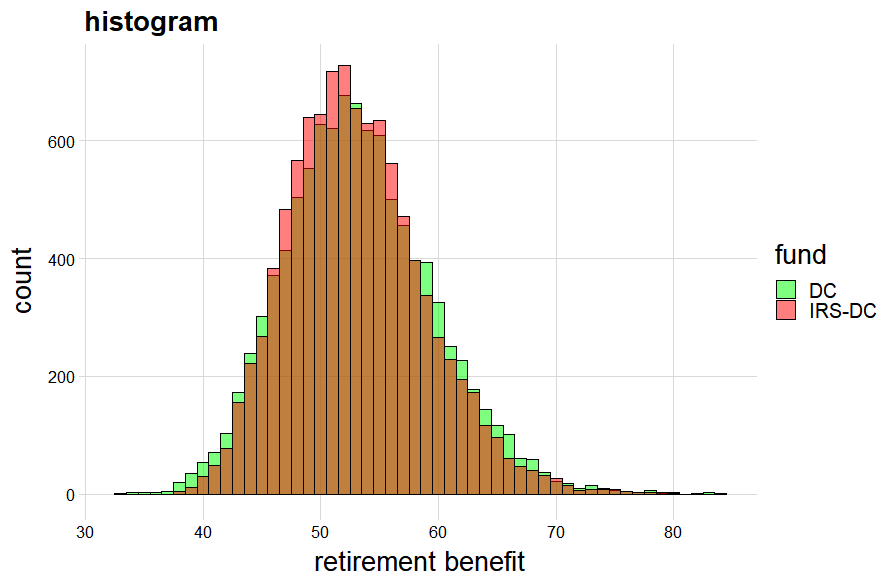}
\vspace{0mm}
\caption{Market 3: $\gamma=5$}
\end{subfigure}%
\caption{ 
Histograms of the retirement benefits (obtained from the 10,000 simulations) of the IRS-DC and pure DC accounts for the generation $i = 41$, for the three market settings and risk aversion $\gamma = 3, 5$.  In each figure, the red and green histograms are those of the IRS-DC and pure DC benefits, respectively; the brown part shows the overlap between the two histograms. 
}
\label{fig:dist-benefits-gamma3-5}
\end{figure}

Figure \ref{fig:dist-benefits-gamma3-5} shows the histograms of the retirement benefits of the IRS-DC and pure DC plan participants for two settings of the risk aversion, $\gamma = 3, 5$; see Section \ref{sec:distribution-benefits} for details.

\subsection{Welfare of Participants} 
\label{sec:appendix-welfare}

\begin{figure}
\centering
\begin{subfigure}[b]{0.55\textwidth}
\includegraphics[width=0.9\linewidth]{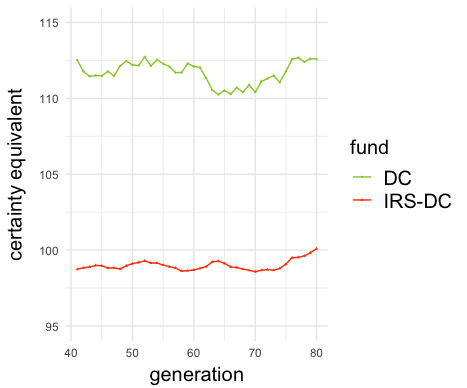}
\vspace{0mm}
\caption{Market 1: $\gamma=3$}
\end{subfigure}%
\begin{subfigure}[b]{0.55\textwidth}
\includegraphics[width=0.9\linewidth]{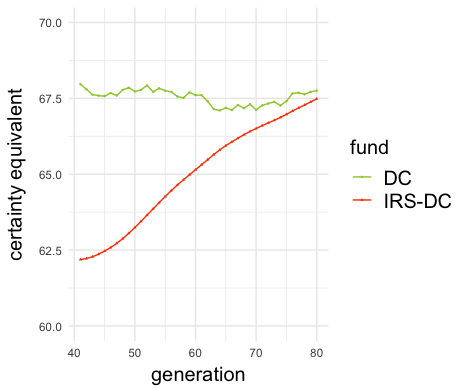}
\vspace{0mm}
\caption{Market 2: $\gamma=3$}
\end{subfigure}%

\vspace{5mm}
\begin{subfigure}[b]{0.55\textwidth}
\includegraphics[width=0.9\linewidth]{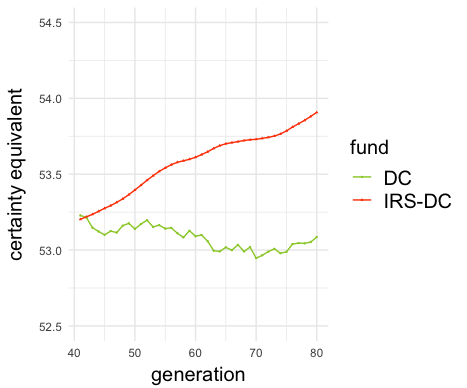}
\vspace{0mm}
\caption{Market 3: $\gamma=3$}
\end{subfigure}%
\begin{subfigure}[b]{0.55\textwidth}
\includegraphics[width=0.9\linewidth]{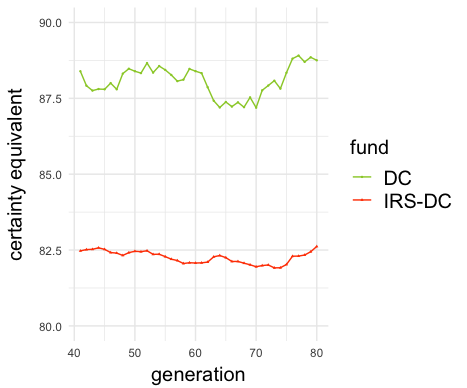}
\vspace{0mm}
\caption{Market 1: $\gamma=5$}
\end{subfigure}%

\vspace{5mm}
\begin{subfigure}[b]{0.55\textwidth}
\includegraphics[width=0.9\linewidth]{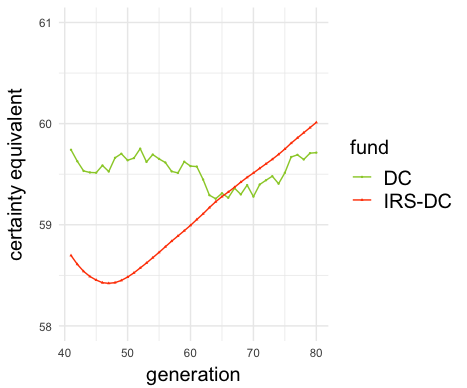}
\vspace{0mm}
\caption{Market 2: $\gamma=5$}
\end{subfigure}%
\begin{subfigure}[b]{0.55\textwidth}
\includegraphics[width=0.9\linewidth]{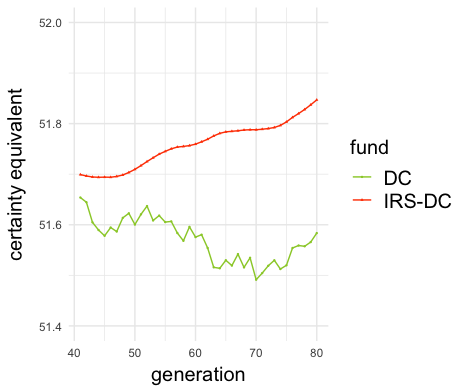}
\vspace{0mm}
\caption{Market 3: $\gamma=5$}
\end{subfigure}%
\caption{
Certainty equivalents of the participants in the IRS-DC and pure DC plans for the generations $i = 41, \dots, 80$, for the three market settings and risk aversion $\gamma = 3, 5$.
}
\label{fig:certainty-equiv-gamma3-5}
\end{figure}

Figure \ref{fig:certainty-equiv-gamma3-5} shows the certainty equivalents of the IRS-DC and pure DC participants for the generations $i = 41, \dots, 80$ for two settings of the risk aversion, $\gamma = 3, 5$; see Section \ref{sec:experiment-CE} for details.

{\color{black}{\section{
Additional Numerical Analyses
}\label{sec:sec6}}}

\textcolor{black}{
We report the results of additional numerical experiments on a path-dependent stochastic investment strategy in Appendix \ref{sec:sec61} and on the calibration of the population structure in Appendix \ref{sec:sec62}. 
}


\subsection{Time-varying Investment Strategy}\label{sec:sec61}

\textcolor{black}{
In the main body of the paper, we consider a constant-mixed strategy that invests a constant fraction $\pi \in (0,1)$ of its asset in the stock for the IRS-DC fund. Here we relax this assumption by considering a time-varying investment strategy $\pi(t) \in (0,1)$ that continuously changes with time $t$ according to the fund's investment performance.}

\textcolor{black}{
Specifically, we consider the funding-ratio-linked investment strategy studied in \citet[Eq.(4)]{goecke2013pension}.
For constants $ \pi_0 \in (0,1)$ and $a \geq 0$, the fraction $\pi(t)$ to be invested in the stock at time $t \geq 0$ is defined as
\begin{equation} \label{eq:pit-2017}
\pi(t) :=  \pi_0 + \frac{a}{\sigma} \ln \left( \frac{A(t)}{L(t)} \right),
\end{equation}
where $A(t)$ and $L(t)$ are the fund's asset and liability, respectively. We set $\pi(t) = 0$ if $\pi(t) < 0$ and $\pi(t) = 1$ if $\pi(t) > 1$.
In this case, the indexation rate $g(t)$ becomes 
$$
g(t) := \mu(t)+\theta \ln \left( \frac{A(t)}{L(t)} \right), \quad 
\text{where} \quad  
 \mu(t) := (\mu-r) \pi(t) +  r -\frac{1}{2} \sigma^2 \pi(t)^2.
$$
Note that $\sigma(t) :=  \sigma \pi(t)$ represents the volatility of the fund's asset. 
}

    

\begin{figure}
\captionsetup{font=normal}
\centering
\begin{subfigure}[b]{0.5\textwidth}
\centering
\includegraphics[width=.8\linewidth]{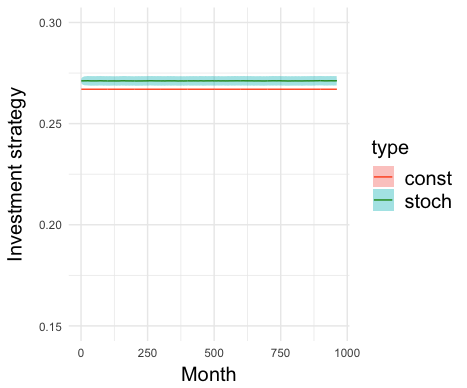}
\vspace{0mm}
\caption{Market 1, $\gamma=10$.}
\end{subfigure}%
\begin{subfigure}[b]{0.5\textwidth}
\centering
\includegraphics[width=.8\linewidth]{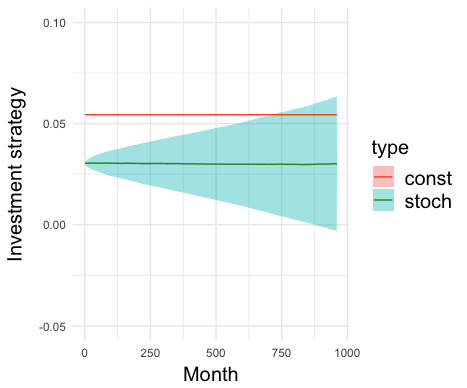}
\vspace{0mm}
\caption{Market 3, $\gamma=10$.}
\end{subfigure}%

\begin{subfigure}[b]{0.5\textwidth}
\centering
\includegraphics[width=.8\linewidth]{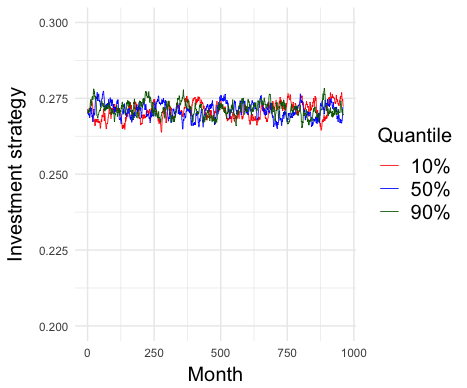}
\vspace{0mm}
\caption{Market 1, $\gamma=10$.}
\end{subfigure}%
\begin{subfigure}[b]{0.5\textwidth}
\centering
\includegraphics[width=.8\linewidth]{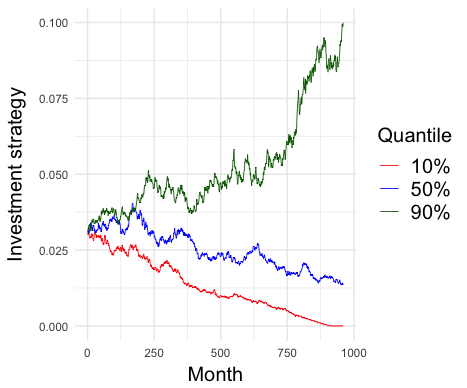}
\vspace{0mm}
\caption{Market 3, $\gamma=10$.}
\end{subfigure}%
\caption{
\textcolor{black}{
In Figures (a) and (b), the green line and confidence band describe the mean and standard deviation of the time-varying investment strategy $\pi(t)$ in \eqref{eq:pit-2017}; the red line indicates the corresponding constant-mix strategy $\pi^*$ (see Table \ref{table:opdesign}).  Figures (c) and (d) plot the paths of $\pi(t)$ for three representative scenarios defined in the same way as Section \ref{sec:funding-ratio-exp}. 
} 
}
\label{fig:strategy_experiment}
\end{figure}

\begin{figure}
\captionsetup{font=normal}
\centering
\begin{subfigure}[b]{0.55\textwidth}
\centering
\includegraphics[width=.8\linewidth]{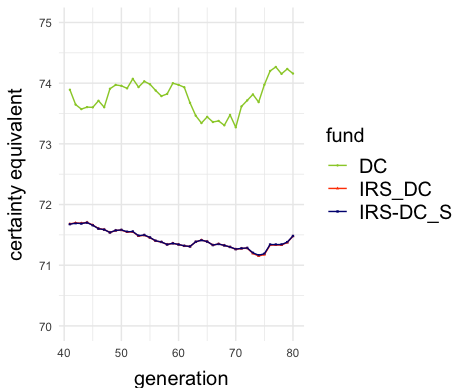}
\vspace{0mm}
\caption{Market 1, $\gamma=10$.}
\end{subfigure}%
\begin{subfigure}[b]{0.55\textwidth}
\centering
\includegraphics[width=.8\linewidth]{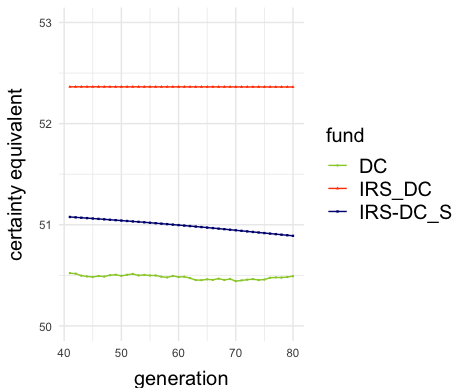}
\vspace{0mm}
\caption{Market 3, $\gamma=10$.}
\end{subfigure}%
\caption{
\textcolor{black}{
Certainty equivalents of generations 41 to 80 obtained with the time-varying investment strategy (IRS-DC-S; blue), the constant-mix strategy (IRS-DC; red) and the life-cycle strategy (DC; green).  In Figure (a), the blue and red curves overlap. 
}
}
\label{fig:strategy_experiment_ce}
\end{figure}

\textcolor{black}{
We optimize the parameters $\pi_0$, $a$ and $\theta$ using Bayesian optimization (where the range of each parameter is [0,1] and the number of iterations is 100), focusing on the risk aversion $\gamma = 10$ in Market 1 (high Sharpe ratio) and  Market 3 (low Sharpe ratio). The results are:  
\begin{align*}
    \text{Market 1:}&\quad\pi_0^{*}=0.2711;\quad a^{*}=0.0118\quad \theta^{*}=0.9995,\\
    \text{Market 3:}&\quad\pi_0^{*}=0.0304;\quad a^{*}=0.0669\quad \theta^{*}=0.0001.
\end{align*}
Figure \ref{fig:strategy_experiment} describes the mean and standard deviation of $\pi(t)$ over 10,000 simulations as a function of $t$, as well as the corresponding constant mixed strategy $\pi^*$. It also shows the paths of $\pi(t)$ for three representative scenarios defined in the same way as Section \ref{sec:funding-ratio-exp}. Figure \ref{fig:strategy_experiment_ce} shows the certainty equivalents of the IRS-DC participants obtained with the time-dependent and constant-mix strategies.
}


\textcolor{black}{ 
Our main findings are as follows: 
}
\textcolor{black}{
\begin{enumerate}
\item 
The investment strategy $\pi(t)$ and (thus the asset volatility $\sigma(t) = \sigma \pi(t)$) appear to be mean-reverting, indicating the existence of an implicit target asset volatility. The standard deviation of $\pi(t)$ is relatively small for Market 1, where $\theta^*$ is large, and is increasing with time $t$ for Market 3, where $\theta^*$ is small. These observations may be explained by the fact that $\pi(t)$ is linked to the log funding ratio $\ln (A(t) / L(t))$, whose volatility decreases for a larger adjustment parameter $\theta$, as analyzed in Sections \ref{sec:dynamics-fund-ratio} and \ref{sec:funding-ratio-exp}.
\item 
Regarding certainty equivalents, the time-dependent investment strategy does not improve upon the constant-mix strategy. One potential reason is that the underlying financial market is too simple, so the constant-mix strategy is sufficient to achieve optimal results. 
Similar investment strategies, such as target-volatility strategies, have been shown to improve post-retirement benefits for pooled annuitants compared to the constant-mixed strategy \citep{li2022managed,olivieri2022target}. It will be interesting to investigate conditions under which the time-dependent investment strategy improves upon the constant-mix strategy for the IRS-DC model.    
\end{enumerate}}

\subsection{Influence of the Population Structure} 
\label{sec:sec62}

\begin{figure}
    \centering
    \includegraphics[width=.5\linewidth]{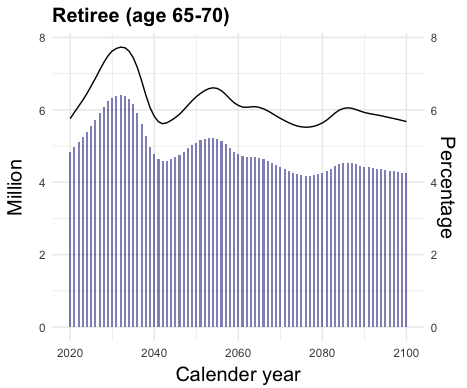}
    \caption{\textcolor{black}{
    Population projection data of Germany for the age group from 65 to 70 in the period between 2021 to 2100.  The black curve represents the percentage of this age group in the total population. 
    }}
    \label{fig:my_retiree}
\end{figure}

\textcolor{black}{ 
In the main body of the paper, we assume that each generation consists of one hypothetical participant; this implicitly assumes that different generations have the same population size. Here, we examine the influence of the population structure on the retirement benefits and certainty equivalents of different generations. To this end, we use the projection data\footnote{The data is available on \href{https://www.populationpyramid.net/}{https://www.populationpyramid.net/}.}  of the German population from 2021 to 2100 to calibrate the population sizes of the  80 generations in the IRS-DC fund. We choose this period to represent an ageing society.    This projection data incorporates the mortality risk, as the population of each generation changes over time. Figure \ref{fig:my_retiree} describes the population projections of the age group from 65 to 70 from 2021 to 2100.  
}

\begin{figure}
\captionsetup{font=normal}
\begin{subfigure}[b]{0.55\textwidth}
\centering
\includegraphics[width=.9\linewidth]{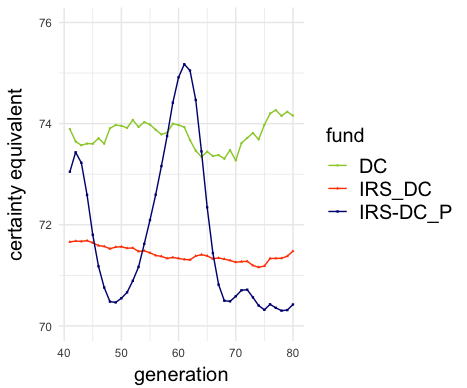}
\vspace{0mm}
\caption{Market 1.}
\end{subfigure}%
\begin{subfigure}[b]{0.55\textwidth}
\centering
\includegraphics[width=.9\linewidth]{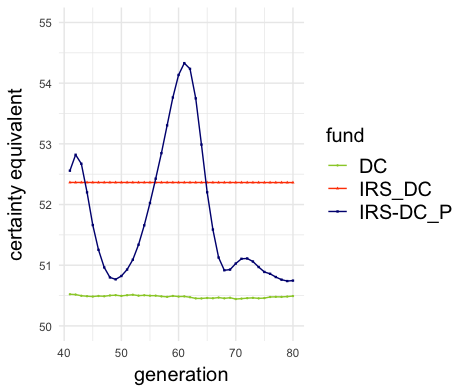}
\vspace{0mm}
\caption{Market 3}
\end{subfigure}%
\caption{\textcolor{black}{
The certainty equivalents of different generations. The green curve (DC) denotes those of the pure DC participants using the life-cycle strategy, the red curve (IRS-DC) those of the IRS-DC with the equal population structure, and the blue curve (IRS-DC-P) those of the IRS-DC with the German population structure. 
}}
\label{fig:population_experiment_ce}
\end{figure}

\textcolor{black}{
In the same way as the main body, we optimize the investment strategy $\pi$ and the adjustment parameter $\theta$ using Bayesian optimization in Markets 1 and 3 with the risk aversion $\gamma = 10$; the results are:
    \begin{align*}
    \text{Market 1:}&\quad\pi^{*}=0.2677;\quad \theta^{*}=0.9999,\\
    \text{Market 3:}&\quad\pi^{*}=0.0356;\quad \theta^{*}=0.0023.
\end{align*}
Figure \ref{fig:population_experiment_ce} describes the certainty equivalents of different generations. One can see that the certainty equivalents change non-smoothly over the generations, compared to those of the IRS-DC fund with the equal population structure.   This result implies that the population structure influences the welfare of the IRS-DC participants and can cause unfairness between different generations. To address this, one could modify the objective function \eqref{problem} or the indexation rate \eqref{benefit-growth-rate} to enforce  fairness among generations. We leave this topic for future research. 
}

   
\end{document}